\documentclass[10pt,journal,compsoc,final]{IEEEtran}

\usepackage[T1]{fontenc}                    
\usepackage[utf8]{inputenc}                 
\DeclareUnicodeCharacter{200B}{{\hskip 0pt}}
\usepackage[english]{babel}

\usepackage[pdftex]{graphicx}               
\usepackage{listings}                       
\usepackage[nolist]{acronym}                
\usepackage[colorlinks,hidelinks]{hyperref}           
\usepackage[all]{hypcap}                    
\usepackage{microtype}                      
\usepackage{amsmath}
\usepackage{amssymb}
\usepackage{pgfplots}
\usepackage{physics}
\usepackage{siunitx}
\usepackage{tikz}
\usepackage{xparse}
\usepackage[capitalise,noabbrev]{cleveref}  
\usepackage{todonotes}
\usepackage[frozencache]{minted}
\usepackage{enumitem}

\usepackage[all]{nowidow}
\newcommand{\term}[1]{%
    \textsc{#1}%
}

\usepackage{import}

\newcommand{\tensor}[1]{%
    \mathcal{#1}%
}
\usepackage{glossaries}
\setacronymstyle{short-long}
\newacronym{GPGPU}{GPGPU}{General-purpose computing on graphics processing units}
\newacronym{GPU}{GPU}{Graphics Processing Unit}
\newacronym{CPU}{CPU}{Central Processing Unit}
\newacronym{CUDA}{CUDA}{Compute Unified Device Architecture}
\newacronym{SM}{SM}{Streaming Multiprocessor}
\newacronym{CTA}{CTA}{Cooperative Thread Array}
\newacronym{SIMT}{SIMT}{Single Instruction Multiple Thread}
\newacronym{JIT}{JIT}{Just-In-Time}
\newacronym{AST}{AST}{Abstract Syntax Tree}
\newacronym{IR}{IR}{Intermediate Representation}
\newacronym{SSA}{SSA}{Static Single Assignment}
\newacronym{PTX}{PTX}{Parallel Thread Execution}
\newacronym{SASS}{SASS}{Streaming Assembler}
\newacronym{ISA}{ISA}{Instruction Set Architecture}
\newacronym{FP64}{FP64}{Double Precision Floating Point}
\newacronym{FP32}{FP32}{Single Precision Floating Point}
\newacronym{FP16}{FP16}{Half Precision Floating Point}
\newacronym{FPU}{FPU}{Floating Point Unit}
\newacronym{MAC}{MAC}{Multiply Accumulate}
\newacronym{HPC}{HPC}{High Performance Computing}
\newacronym{ML}{ML}{Machine Learning}
\newacronym{DL}{DL}{Deep Learning}
\newacronym{GEMM}{GEMM}{General Matrix Multiplication}
\newacronym{API}{API}{Application Programming Interface}
\newacronym{BLAS}{BLAS}{Basic Linear Algebra Subprograms}
\newacronym{DLA}{DLA}{Dense Linear Algebra}
\newacronym{HLSL}{HLSL}{High Level Shading Language}
\newacronym{OpenGL}{OpenGL}{Open Graphics Library}
\newacronym{OpenCL}{OpenCL}{Open Computing Language}
\newacronym{TTGT}{TTGT}{Transpose-Transpose-GEMM-Transpose}
\newacronym{LoG}{LoG}{Loop-over-GEMMs}
\newacronym{GETT}{GETT}{GEMM-like Tensor-Tensor contraction}
\newacronym{MKL}{MKL}{Math Kernel Library}
\newacronym{ATLAS}{ATLAS}{Automatically Tuned Linear Algebra Software}
\newacronym{BLIS}{BLIS}{BLAS-like Library Instantiation Software}
\newacronym{TBLIS}{TBLIS}{Tensor BLIS}
\newacronym{MAGMA}{MAGMA}{Matrix Algebra on GPU and Multicore Architectures}
\newacronym{CUTLASS}{CUTLASS}{CUDA Templates for Linear Algebra Subroutines}
\newacronym{WMMA}{WMMA}{Warp Matrix Multiply Accumulate}
\newacronym{BF16}{BF16}{Bfloat16}
\newacronym{TF32}{TF32}{TensorFloat-32}
\newacronym{GTC}{GTC}{GPU Technology Conference}
\newacronym{PL}{PL}{Programming Language}
\newacronym{DSL}{DSL}{Domain-Specific Language}
\newacronym{TC}{TC}{Tensor Core}
\newacronym{TCT}{TCT}{Tensor ConTraction}

\pgfplotsset{compat=1.14}

\usepackage[nocompress]{cite}

\usepackage{float}
\newfloat{span}{tbp}{los}

\usepackage{pifont}

\setminted{fontsize=\small,linenos,numbersep=5pt,frame=lines,framesep=2mm,breaklines}
\makeatletter
\def\lst@makecaption{%
  \def\@captype{table}%
  \@makecaption
}
\makeatother

\makeatletter
\let\org@@cref\@cref
\renewcommand*{\@cref}[2]{%
  \edef\process@me{%
    \noexpand\org@@cref{#1}{\zap@space#2 \@empty}%
  }\process@me
}
\makeatother

\usepackage{tikz}

\usepackage{textcomp} 
\usepackage{siunitx}
\begin{document}

%
%
 \IEEEoverridecommandlockouts
   \IEEEpubid{\color{red}\begin{tabular}[t]{@{}l@{}}
     This paper was submitted to IEEE TPDS. \copyright~2021 IEEE. \\
     Personal use of this material is permitted. Permission from IEEE
     must be obtained for all other uses, in any current or  future media, including
     reprinting/republishing this material for \\ advertising or promotional purposes, creating
     new collective works, for resale or redistribution to servers or lists, or reuse of any
     copyrighted component of this work in other works.
   \end{tabular}}

%
%

\lstset{%
  basicstyle          = \ttfamily%
                        \lst@ifdisplaystyle\scriptsize\fi,
  keywordstyle        = \color{lst2},
  stringstyle         = \color{lst1},
  commentstyle        = \itshape\color{lst3},
  showstringspaces    = false,
  frame               = top,
  frame               = bottom,
  framextopmargin     = 2pt,
  framexbottommargin  = 2pt,
  framexleftmargin    = 17pt,
  xleftmargin         = 17pt,
  belowskip           = 0ex,
  numbers             = left,
  numbersep           = 7pt,
  escapechar          = \¶
}

\lstdefinelanguage{CUDAC}[ANSI]{C}{%
  morekeywords={%
    __global__,__device__,%
    threadIdx,blockIdx,blockDim,gridDim},%
}


\lstdefinelanguage{Julia}{%
  morekeywords={%
    type,primitive,abstract,struct,new,%
    function,@generated,macro,module,where,%
    begin,end,do,%
    try,catch,return,%
    const,export,import,using,%
    if,elseif,else,for,while,break,continue,%
    true,false,quote},%
  sensitive=true,%
  alsoother={\$},%
  morecomment=[l]\#,%
  morecomment=[n]{\#=}{=\#},%
  morestring=[s]{"}{"},%
  morestring=[m]{'}{'},%
}[keywords,comments,strings]

\lstdefinelanguage{JuliaCUDA}[]{Julia}{%
  morekeywords={%
    @cuda,%
    CuArray,CuDeviceArray,%
    CuDevice,CuContext,destroy,%
    blockIdx,blockDim,threadIdx,threadDim,gridDim,warpsize},%
}

%
%

\begin{acronym}

\acro{api}[API]{Application Programming Interface}
\acro{ast}[AST]{Abstract Syntax Tree}
\acro{cuda}[CUDA]{Compute Unified Device Architecture}
\acro{dsl}[DSL]{Domain Specific Language}
\acro{ffi}[FFI]{Foreign Function Interface}
\acro{gpu}[GPU]{Graphics Processing Unit}
\acro{cpu}[CPU]{Central Processing Unit}
\acro{ir}[IR]{Intermediate Representation}
\acro{isa}[ISA]{Instruction Set Architecture}
\acro{jit}[JIT]{Just-in-Time}
\acro{llvm}[LLVM]{Low-Level Virtual Machine}
\acro{ptx}[PTX]{Parallel Thread Execution}
\acro{loc}[LOC]{lines of code}
\acro{nvrtc}[NVRTC]{NVIDIA Runtime Compilation}

\end{acronym}

%
%

\hypersetup{pdfauthor={Thomas Faingnaert},
            pdftitle={Flexible Performant GEMM Kernels on GPUs},
            pdfkeywords={Julia, CUDA, WMMA}}

\definecolor{NavyBlue}{cmyk}{0.94,0.54,0,0}
\hypersetup{citecolor = NavyBlue,
            linkcolor = NavyBlue,
            urlcolor = NavyBlue}

\title{Flexible Performant GEMM Kernels on GPUs}

\author{
  Thomas Faingnaert, Tim Besard, Bjorn De Sutter, \IEEEmembership{Member,~IEEE}%
  \IEEEcompsocitemizethanks{\IEEEcompsocthanksitem T. Faingnaert and
                            B. De Sutter are with the Department of
                            Electronics~and~Information~Systems,
                            Ghent~University, Belgium. T. Besard works for Julia Computing.\protect\\
  \texttt{thomas.faingnaert@ugent.be;tim@juliacomputing.com}\protect\\
  Corresponding author: \texttt{bjorn.desutter@ugent.be}}
  \thanks{Manuscript received X, ; revised X.}
}

\markboth{IEEE Transactions on Parallel and Distributed Systems}{Faingnaert \MakeLowercase{\textit{et al.}}}

%

\IEEEtitleabstractindextext{%
\begin{abstract}
General Matrix Multiplication or GEMM kernels take centre place in high performance computing and machine learning. Recent NVIDIA GPUs include GEMM accelerators, such as NVIDIA’s Tensor Cores.  Their exploitation is hampered by the two-language problem: it requires either low-level programming which implies low programmer productivity or using libraries that only offer a limited set of components. Because rephrasing algorithms in terms of established components often introduces overhead, the libraries' lack of flexibility limits the freedom to explore new algorithms. Researchers using GEMMs can hence not enjoy programming productivity, high performance, and research flexibility at once.
In this paper we solve this problem. We present three sets of abstractions and interfaces to program GEMMs within the scientific Julia programming language. The interfaces and abstractions are co-designed for researchers' needs and Julia's features to achieve sufficient separation of concerns and flexibility to easily extend basic GEMMs in many different ways without paying a performance price. Comparing our GEMMs to state-of-the-art libraries cuBLAS and CUTLASS, we demonstrate that our performance is in the same ballpark of the libraries, and in some cases even exceeds it, without having to write a single line of code in CUDA C++ or assembly, and without facing flexibility limitations.

%

%

%



%

\end{abstract}

\acresetall

}

\maketitle

\section{Introduction}
\label{sec:introduction}




\gls{GEMM} kernels form the core of many computations in the fields of \gls{HPC} and \gls{ML}. In \gls{HPC}, GEMM is at the core of linear algebra~\cite{blas2017home}, including dense linear algebra~\cite{abdelfattah2019towards,haidar2018harnessing-iterative-refinement}, and is used for earthquake simulation~\cite{ichimura2018fast}, plasma visualisation~\cite{haidar2018harnessing-hpc-scientific-applications}, and weather and climate prediction~\cite{mehta2019getting}. In \gls{ML} they are used to train neural networks including fully connected layers in traditional neural networks, convolutional neural networks, long short-term memory cells, and natural language processing~\cite{yan2020demystifying,nvidia2020deep}. To accelerate their computations, researchers in the mentioned domains have relied on the massively parallel computing resources of \glspl{GPU}.

To answer the demand for more efficient \glspl{GEMM}, recent \glspl{GPU} include matrix multiplication accelerators, such as NVIDIA's \glspl{TC}~\cite{nvidia2020v100}. Researchers can exploit these resources in two ways. They can express their algorithms in high-level \glspl{PL} such as Python and express them in terms of established \gls{GEMM} variants for which efficient implementations are available in third-party libraries such as \term{cuBLAS} or \term{CUTLASS}. This approach offers high research productivity, at the cost of being limited to the \glspl{API} and \gls{GEMM} implementations available in the libraries. In many domains, this lack of flexibility is problematic. When non-standard, more generalised \glspl{GEMM} as needed in neural networks~\cite{barham2019machine}, convolutional networks~\cite{nvidia2020deep}, fluid dynamics~\cite{rink2018cfdlang}, electromechanics~\cite{poya2017ahighperformance}, computational chemistry~\cite{auer2006automatic}, or any other computation on multidimensional tensors~\cite{springer2017landscape,nelson2015generating,springer2018design,napoli2014towards,li2015aninput,solomonik2013cyclops,bader2006algorithm,kim2019acode,matthews2018high,psarras2021landscape} are rephrased in terms of standard \gls{GEMM} kernels available in libraries, additional custom kernels need to be launched in between the \gls{GEMM} kernels for things such as precision conversions, layout conversions (transpositions), type conversions, bias operations, element-wise operations, etc. These extra kernels introduce huge overheads because they have a massive impact on the traffic to the very slow global memory.

Alternatively, researchers can rewrite the most demanding parts of their software in lower-level \glspl{PL} such as CUDA C/C++~\cite{nvidia2020cuda} or OpenCL~\cite{khronos2020opencl}. This decreases their productivity, however, and they now require much more \gls{PL} and \gls{GPU} programming model knowledge outside their own application domain. In short, many researchers working with \gls{GEMM}-like algorithms suffer from the \emph{two-language problem}. They cannot achieve high performance, high research productivity, and algorithmic flexibility together. 

The scientific \gls{PL} Julia is designed to overcome the two-language problem~\cite{julia2020home}. It offers a high-level syntax, dynamic typing, managed memory, meta-programming, multiple dispatch, and other features that increase programmer productivity. Julia's compiler is based on type inference and just-ahead-of-time compilation, which allows it to generate code devoid of much of the run-time overhead (e.g.\ in the form of dynamic type checks) that other \glspl{PL} pay for supporting the mentioned features. On \glspl{CPU}, Julia code is comparable in performance to C, C++, and Fortran code~\cite{julia2020benchmarks}. Through the \term{CUDA.jl} package, it is possible to program NVIDIA GPUs directly in Julia, at a high abstraction level of arrays or at the lower-level of CUDA-like kernels~\cite{besard2019effective,besard2019rapid}. Before our research, \glspl{TC} were not supported in \term{CUDA.jl}, however. The package and its high-level \glspl{API} were hence of limited use to many researchers.

In our research, we set out to overcome this issue in three steps. First, we developed support for \glspl{TC} in the Julia compiler and libraries through a WMMA API of wrapper functions around so-called compiler intrinsics. This allows for exploiting \glspl{TC} in kernels written with the lower-level support in \term{CUDA.jl}. This low-level API only focuses on the WMMA operation. It does not free the programmer from the cumbersome task of coordinating the memory traffic in the memory hierarchy to move data to and from the \glspl{TC}. So secondly, we developed a tiling API in Julia that allows programmers to coordinate the memory traffic to and from \glspl{TC} at a high abstraction level, and, importantly, without paying a price in terms of performance. The low-level API and the tiling API enable efficient use of \glspl{TC} and the \gls{GPU} memory hierarchy, but to reach good performance, the \gls{GEMM} computations themselves, possibly with fused additional computations, then still need to be programmed at a rather low level of abstraction requiring a lot of expertise. In the final step, we therefore developed a high-level \gls{GEMM} API in Julia that allows programmers to express and combine a range of extensions of basic \gls{GEMM} computations in an abstract, intuitive way, without having to pay an unacceptable price in performance. Combined, these three APIs solve the two-language problem to a great extent with respect to hardware resources such as \glspl{TC}.

Our main result is that we get performance in the same ballpark of hand-tuned libraries and in some cases even much better performance, without having to write a single line of code in a lower-level \gls{PL} and without being limited to the specific \gls{GEMM} versions supported by the libraries.

This paper focuses on the tiling and \gls{GEMM} APIs. After providing the necessary background in Section~\ref{sec:background}, Section~\ref{sec:tiling} discusses requirements for a tiling API, presents our novel way for abstracting tiling and the Julia API we designed based on that abstraction, and demonstrates and evaluates the API on a number of stages in \gls{GEMM} computations. Section~\ref{sec:matmul} discusses the requirements for flexibility in \glspl{GEMM} in more detail. We present the different building blocks at the basis of our Julia \gls{GEMM} API that provide that flexibility in an intuitive manner, and we demonstrate the API on a number of examples. In Section~\ref{sec:evaluation}, our final contribution is a performance evaluation of multiple variants of \gls{GEMM} computations, showing that we get relatively close to the performance of hand-tuned libraries like \term{cuBLAS}, \term{CUTLASS}, and \term{cuTENSOR} without having to write any single line in a lower-level \gls{PL}. The paper then ends with a discussion of some related work in Section~\ref{sec:related}, the availability of our artefacts in Section~\ref{sec:availability}, and with a conclusion and a look forward in Section~\ref{sec:conclusions}.

\section{Background}
\label{sec:background}

\subsection{GPU programming}%
\label{sub:gpu_programming}

The main difference between programming GPUs versus CPUs is their underlying programming model.
GPUs are massively parallel processors, meaning that a large number of threads execute the same function in parallel.
In GPU parlance, this function is commonly referred to as a \emph{kernel}.

GPU threads are organised in a thread hierarchy~\cite{nvidia2020cuda}.
Since our main interest is in NVIDIA GPUs, we limit our discussion to NVIDIA's CUDA programming model.
\emph{Threads} are the smallest unit of execution in the hierarchy.
The hardware groups them into sets of 32 threads called \emph{warps}.
Threads in the same warp execute in a SIMT (Single Instruction Multiple Thread) fashion.
These threads must hence execute the same instruction at the same time, possibly on different data.
Threads are also grouped by the programmer into \emph{blocks}.
Threads in the same block can communicate efficiently, so that they can cooperate on a common task.
Finally, the set of all blocks on the GPU device is called the \emph{grid}.

Similarly to threads, GPU memory is also ordered hierarchically.
We are mainly interested in three parts of this hierarchy, which correspond directly to levels in the thread hierarchy.
The \emph{register} file is the fastest type of memory.
Each thread typically has access to 255 registers. Each block has its own set of \emph{shared memory}, that may be used by threads in the same block to communicate.
Finally, \emph{global memory} can be accessed by all threads on the device, regardless of which block they belong to.
Global memory has the largest capacity, but also has much higher latency and lower throughput.

To fully exploit the available resources on a GPU, programmers can either use low-level \glspl{PL} like CUDA C/C++~\cite{nvidia2020cuda} or OpenCL~\cite{khronos2020opencl} to program their own kernels, or they can use the foreign function interface of high-level \glspl{PL} such as Python to invoke kernels in libraries. The former option requires quite some knowledge in GPU programming models, forces the programmers to write quite some boilerplate code to manage data in memories and configure the kernels, and offers little performance portability, so manual (re)tuning of code is necessary when porting the code to different devices. Popular libraries such as \term{cuBLAS} contain kernel versions tuned for many different devices to overcome the performance portability issue.

\subsection{Julia Programming Language}%
\label{sub:julia}

The open-source \gls{PL} Julia features a high-level syntax~\cite{julia2020docs}.
A central paradigm in its design is the way it handles dispatch, the process by which the compiler chooses which implementation of a function to use for a given function call.
Julia uses a \emph{multiple dispatch} scheme, which means that this choice depends on the types of \emph{all} of a function's arguments.

Julia's type system is \emph{dynamic}, meaning that the types of expressions are not necessarily known statically.
However, Julia inherits some of the advantages of static type systems through several features of its compiler.
For one, the Julia compiler applies type inference to deduce the types of values used by the program.
Code is then \emph{specialised} based on this information, e.g.\ function calls are devirtualised, dynamic type checks are removed, etc.
This style of compilation, dubbed \emph{just-ahead-of-time}, has the performance of ahead-of-time compiled \glspl{PL} with the flexibility of a just-in-time compiled one.
We rely on this design to seamlessly compose a \gls{GEMM} computation from all involved components, i.e.\ beyond what normal layering of libraries at different layers of abstraction allows as is typically done with other \glspl{PL}.

Julia's compiler is built on top of LLVM, a compiler infrastructure project commonly used in research and industry~\cite{llvm2020home}.
Julia's compilation process consists of a couple steps.
First, Julia code is converted to an IR (Intermediate Representation) that is used for type inference, called Julia IR.
Next, Julia IR is lowered to LLVM IR, the representation that LLVM uses.
From this point onwards, the LLVM framework takes control of the compilation process.
LLVM contains a set of backends, one for each target architecture that LLVM supports.
The backend corresponding to the current architecture will then convert this LLVM IR to native instructions.

The Julia package \term{CUDA.jl} reuses part of the Julia compilation process to allow executing kernels written in Julia on NVIDIA GPUs~\cite{besard2019effective}.
In particular, the aforementioned compilation pipeline is run to the point where Julia IR is lowered to LLVM IR.
The LLVM IR is intercepted and sent to the LLVM NVPTX backend instead of the backend of the host architecture.
This NVPTX backend converts the IR to \gls{PTX} instructions, the virtual instruction set of NVIDIA GPUs.

With \term{CUDA.jl}, it is possible to program NVIDIA GPUs at the lower-level of CUDA-like kernels and at the higher abstraction level of arrays~\cite{besard2019effective}. The former involves less boilerplate and verbosity, and makes reusing code easier compared to programming in CUDA C/C++. The latter enables much more productive programming~\cite{besard2019rapid}.

Julia's multiple dispatch enables transparent exploitation of performance-optimised functionality from popular GPU libraries. For example, for any function from the \term{cuBLAS} library, a Julia package can contain a generic implementation in pure Julia code that operates for all (numeric) data types and that hence accepts all arguments of type \mintinline{text}{Number}. In addition, the package can contain wrappers that each only accept a more concrete argument type such as \mintinline{text}{Float32} and that invoke the corresponding \term{cuBLAS} function for that type. Users of the package can then invoke the function on any type they want. If it is supported by the \term{cuBLAS} library, they will get optimal performance ``for free''.

\subsection{Tensor Cores}%
\label{sub:tensor_cores}
Each \gls{TC} performs a matrix multiply-accumulate expression of the form \( D = A \cdot B + C \).
\glspl{TC} support a limited set of possible data types for these matrices.
For example, if the \( A \) and \( B \) matrices are stored as 16-bit floating point values, the \( C \) and \( D \) matrices are 32-bit floating point.

NVIDIA exposes \glspl{TC} in C++ in the so-called \gls{WMMA} API.
WMMA instructions must be used by all threads in a warp in a SIMT fashion.
Each thread that cooperates in a warp-wide WMMA operation holds a part of each matrix in its registers, called a \emph{fragment}.
In the remainder of this paper, unless stated differently, we will assume that \( A \) is an \( M \times K \) matrix, \( B \) is a \( K \times N \) matrix, and \( C \) and \( D \) are \( M \times N \) matrices.
The tuple \( (M, N, K) \) is called the \emph{shape} of the WMMA operation.
Not all possible values of \( M \), \( N \), and \( K \) are allowed, as WMMA restricts the set of possible shapes. Conceptually, WMMA consists of three separate steps:
\begin{enumerate}
    \item Load the input matrices \( A \), \( B \), and \( C \) from memory into WMMA fragments using a WMMA \term{load} operation.
    \item Perform the matrix multiply-accumulate using a WMMA \term{mma} operation, resulting in a fragment of \( D \).
    \item Store the resultant \( D \) fragment to memory using a WMMA \term{store} operation.
\end{enumerate}
In CUDA C++ each step corresponds to an overloaded C++ function.
Calls to these functions are mapped one-to-one onto the corresponding WMMA \gls{PTX} instruction by the compiler.

To add support for WMMA to \term{CUDA.jl}, we reused the pre-existing WMMA \gls{PTX} intrinsics in the NVPTX backend.
This necessitated adaptations to Julia's compiler, in particular to the code generation process.
Our WMMA API consists of two different layers.
The lowest layer consists of Julia wrapper functions that are mapped one-to-one to these intrinsics.
The second layer is a high-level interface, similar to CUDA C++'s version of WMMA.
It consists of \mintinline{text}{load_a}, \mintinline{text}{load_b}, \mintinline{text}{load_c}, \mintinline{text}{mma}, and \mintinline{text}{store_d} functions, which call the intrinsic wrapper corresponding to the argument types.

At its launch with the Volta architecture in 2017, WMMA only supported \( 16 \times 16 \times 16 \) multiply-accumulates of FP16 matrices.
More recent GPU architectures extend the interface with new data types and shapes.
Turing's second generation \glspl{TC}, introduced in 2018, add support for 8-bit, 4-bit, and 1-bit data types, along with new WMMA shapes depending on the data type used.
The most recent version of WMMA includes support for FP64, bfloat16, and TF32 data types, and was launched in May 2020 with the introduction of Ampere.

\glspl{TC} can also be used through libraries instead of WMMA. NVIDIA's \term{cuDNN} library contains \gls{TC} kernels for common \gls{ML} algorithms.
\gls{ML} frameworks such as TensorFlow, PyTorch, and MXNet use \term{cuDNN} for training and inference.
\term{cuBLAS}, \term{cuBLASLt}, and \term{CUTLASS} contain optimised GEMM kernels for \gls{HPC} applications.
NVIDIA's \term{cuTENSOR} builds on \term{CUTLASS} and contains Tensor-Core-accelerated kernels for tensor computations.

\section{Abstractions for recursive blocking}%
\label{sec:tiling}

\subsection{Requirements}
Matrix multiplication is rich in data reuse. For example, multiplying square matrices of size \( N \) requires \( \mathcal{O}(N^3) \) floating point operations, but only \( \mathcal{O}(N^2) \) storage, so each element is reused roughly \( \mathcal{O}(N) \) times. To exploit this reuse, data needs to be re-accessed as much as possible in faster memories. When all data does not fit into the fastest memories, the transfers between different memories in the hierarchy need to be coordinated carefully to maximise reuse.

For GEMMs, the general idea is to copy tiles of the input matrices up into the memory hierarchy: from global memory to shared memory and from there to registers. The size of the tiles in each step is chosen such that they fit in the available memory. As computations of different tiles of the resultant matrix are independent, those can be performed completely in parallel to maximise the resource utilisation of massively parallel GPUs. Because of the one-to-one mapping between levels of threads and the memory hierarchy, each of the tiled copy operations is also performed cooperatively, by all threads in the relevant part of the thread hierarchy.

Consider again the GEMM of \( D = A \cdot B + C \). With tiling, this GEMM will consist of the following stages:
\begin{enumerate}
    \item Copy a tile of \( C \) from global memory to shared memory, cooperatively by all threads in a block.
    \item Copy a tile of \( C \) from shared memory to registers, cooperatively by all threads in a warp.
    \item Iterate over dimension \( K \), stride $=$ block tiling size.
        \begin{enumerate}
            \item Copy a tile of \( A \) from global memory to shared memory, cooperatively by all threads in a block.
            \item Do the same for a tile of \( B \).
            \item Iterate over dimension \( K \), stride $=$ warp tiling size.
                \begin{enumerate}
                    \item Copy a tile of \( A \) from shared memory to registers, cooperatively by all threads in a warp.
                    \item Do the same for a tile of \( B \).
                    \item Compute a tile of \( D \), given the \( A \), \( B \), and \( C \) tiles, cooperatively by all threads in a warp.
                \end{enumerate}
        \end{enumerate}
    \item Copy a tile of \( D \) from registers to shared memory, cooperatively by all threads in a warp.
    \item Copy a tile of \( D \) from shared memory to global memory, cooperatively by all threads in a block.
\end{enumerate}

For a WMMA GEMM, stages 2, 3.c.i, and 3.c.ii correspond to WMMA \term{load} operations, stage 3.c.iii to \term{mma} operations, and stage 4 to WMMA \term{store} operations.

This form of recursive blocking is an absolute requirement to achieve good performance, but it is also complex to program. Tile sizes need to be chosen in function of the hardware and the number of dimensions and the sizes of the data, indexing of the matrices depends on their layouts, optimal tiling parameters can differ between memory and computational stages. Determining the tiling parameters is complex, and encoding all address computations in the actual code is cumbersome, error-prone, and results in code that is hard to comprehend, port, and maintain.
To make it easier to program general GEMM computations with recursive blocking, we developed a novel \gls{API} with which the tiling computations can be abstracted to a much higher level. The requirements we put forward for this \gls{API} are the following:
\begin{itemize}
\item \emph{Code readability} to ease writing kernels that use blocking.
\item \emph{Zero performance cost} compared to manually expressed address computations of all tiles.
\item \emph{Support for multiple dimensions} (>2) to support \glspl{TCT} and batched GEMMs.
\item \emph{Recursive blocking} for the different levels of the memory hierarchy through independent tiling parameters.
\item \emph{WMMA-compatibility} to exploit WMMA.
\end{itemize}

\subsection{Abstract Operations}
To meet the requirements, we propose a novel abstraction consisting of four different operations on tiles.

\begin{figure}[t]
  \centering
    \includegraphics[width=\columnwidth]{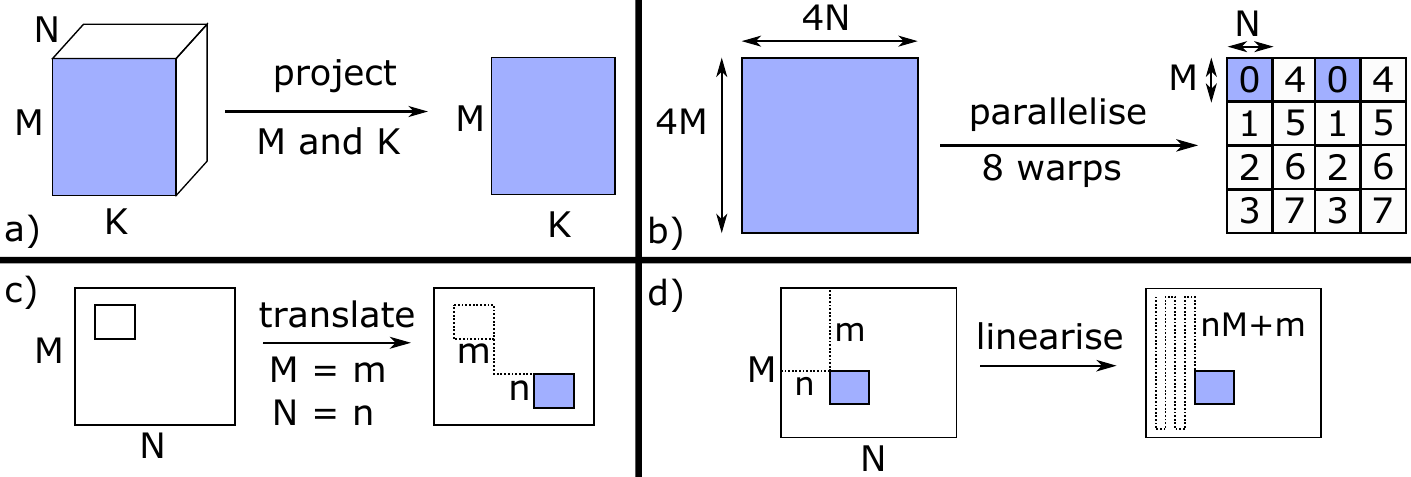}
    \vspace{-0.5cm}
    \caption{Projection, parallelisation, translation and linearisation of tiles.}
    \label{fig:tiling-combined}
\end{figure}

\emph{Projection} is the first abstraction. The compute stages of \gls{GEMM} will use tiles that refer to the three-dimensional iteration space \( (M, N, K) \).
In the memory stages of \gls{GEMM}, we typically only need two of these dimensions.
For example, to load a slice of the \( A \) matrix, we are only interested in the \( M \) and \( K \) dimension.
Projecting a tile reduces its dimensionality by dropping one or more of its dimensions, as shown in \Cref{fig:tiling-combined}a.
The projection abstraction thus allows us to easily reduce the original three-dimensional tile to a tile containing only the relevant dimensions.

\emph{Parallelisation} is the most important operation of the tiling \gls{API}.
It corresponds to the recursive subdivision of tiles in smaller tiles, and the subsequent parallelisation of the resulting subtiles over a set of collaborating entities, such as thread blocks or warps.
Consider the example in \Cref{fig:tiling-combined}b.
A tile of size \( 4M \times 4N \) is divided in subtiles, each of size \( M \times N \).
These subtiles are handled in parallel by a set of 8 cooperating warps, indicated by the numbers 0--7.
The set of all cooperating warps do not need to cover the entire tile. In the example, there are 16 subtiles but only 8 warps.
This means that each warp will handle 2 of these 16 subtiles.
This parallelisation can be applied recursively, by dividing each of these subtiles into sub-subtiles, where each sub-subtile is handled by one thread.


\emph{Translation} moves a tile over a specified distance in each dimension.
In the example of \Cref{fig:tiling-combined}c, a two-dimensional tile is moved over a distance \( m \) in the \( M \) dimension, and a distance \( n \) in the \( N \) dimension.
The translation operation is useful in cases where the reference point of a tile needs to be changed.
For example, consider a tile referring to a submatrix stored in global memory.
The coordinates of this tile are specified relative to the first element in the first row of the parent matrix in global memory.
To copy this submatrix to shared memory, we need to express the tile relative to the first element stored in shared memory, which may be different.
To accomplish this, we can simply translate the tile over the correct distance.


\emph{Linearisation} is used to convert a tile's location from a Cartesian index to a linear index.
This is needed to calculate the offset of a tile in memory, relative to the base pointer of the parent tile.
In the example of \Cref{fig:tiling-combined}d, we consider a subtile at a Cartesian offset of \( (m, n) \) from its parent tile with size \( (M, N) \).
Linearisation results in the linear offset of this tile, relative to the top-left corner of the parent tile.
The linearisation process assumes that the matrix is stored in column major ordering, as this is the convention that Julia uses.
In this case, we need to span \( n \) columns of \( M \) elements each, and an additional \( m \) elements to reach the subtile.
This corresponds to a linear index of \( nM + m \).


\subsection{A tiling \gls{API} for Julia}
\label{sub:tiling_API}
To overcome challenges in developing a concrete \gls{API} based on the four abstractions while meeting all requirements, we relied on some high-level Julia features.

First, a tile is fully determined by its position and its size. Our tiling \gls{API} contains a \mintinline{text}{Tile} struct that stores this information. Storing the size in a field of this struct does, however, not suffice to meet the zero-cost requirement. In Julia, each function is JIT-compiled once for each combination of argument types occurring during the execution of the program. During each such JIT-compilation, no specialisation takes place based on the values of the arguments. In order for the compiler to generate high quality code for the different stages in tiled GEMMs, it needs to know how many registers are needed when transferring slices into registers. In other words, the sizes of the tiles need to be available at compile time, such that specialised code can be generated per tile size. Moreover, no dynamic type checking should be necessary in the generated code. To obtain the required specialisation, yet avoid that any dynamic type checks are needed, we defined \mintinline{text}{Tile} to be a parameterised type, where one of the type parameters (rather than a field) is the size of the tile. Julia's type inference can then obtain all the necessary information to enable specialised code generation in the JIT compiler without running into type instability issues~\cite{julia2020docs}.

Secondly, we observe that when we want to implement \gls{GEMM} using tiling, we typically do not think in terms of the first or second dimension of a tile. Instead, a tile that represents a slice of the \( A \) matrix of size \( M \times K \) has \( M \) and \( K \) dimensions. Rather than writing the position as \mintinline{text}{pos[1]}, we can increase readability by naming the dimensions, so that we may write \mintinline{text}{pos.M}. This form of syntactic sugar can easily be achieved with the existing Julia type \mintinline{text}{NamedTuple}, which we use to store both the position and size of a tile.

Thirdly, we observed a form of structural bias in the Julia-LLVM tool flow with respect to address computations such as those typically occurring in recursive blocking code. The problem is that in many stages of the tiled computation, different threads operate on tiles at different positions in the input matrices or tensors. If the position stored in the \mintinline{text}{Tile} is simply a single position, unnecessarily complex PTX code is generated. However, if the position is split into a thread-dependent base index and a thread-independent offset, the compiler generates code that efficiently exploits the available register + constant addressing mode.

\begin{listing}
\begin{minted}{julia}
struct Tile{size, names, T}
  base::NamedTuple{names, T}
  offset::NamedTuple{names, T}
end
\end{minted}
    \caption{The definition of a Tile in the Julia tiling \gls{API}.}
    \label{lst:tiling_tile}
\end{listing}

The final definition of the parameterised \mintinline{text}{Tile} type is shown in \Cref{lst:tiling_tile}.
With this definition, the implementation for the translate operation is fairly simple.
We define the function \mintinline{text}{translate(tile, dist)} that returns a new tile with the same size and offset, but where the base is the element-wise sum of the original tile's base and the argument \mintinline{text}{dist}.
This essentially moves the multidimensional tile over the distance specified by the argument.

The first argument of \mintinline{text}{linearise(coord, dim)} represents the coordinate of the tile.
We do not take the tile itself as an argument, so that \mintinline{text}{linearise} can be used for both the base and offset of a tile.
Instead of having a separate \mintinline{text}{linearise} function for base and offset, we may simply write \mintinline{text}{linearise(tile.base, ...)} and \mintinline{text}{linearise(tile.offset, ...)}.
The second argument \mintinline{text}{dim} represents the size of the parent tile.
To convert the Cartesian index to a linear index, we use the \mintinline{text}{LinearIndices} type from the Julia standard library.
This way, we can both reuse functionality, and ensure the linearise operation works for any number of dimensions.

One option to project tiles is to define a function \mintinline{text}{project(tile, dims)}, where \mintinline{text}{dims} contains a list of the dimensions to keep.
A projection of a tile to the \( M \) and \( N \) dimension could then be written as \mintinline{text}{project(tile, (:M, :N))}.
We instead opted to use Julia's extensibility.
In Julia, the syntactic construct \mintinline{text}{a.b} is converted to a call to \mintinline{text}{Base.getproperty(a, :b)}~\cite{julia2020docs}.
Through the multiple dispatch mechanism, we override this function such that one can express the project operation as \mintinline{text}{tile.MN} instead of \mintinline{text}{project(tile, (:M, :N))}.

\Cref{lst:tiling_project} shows part of the implementation of the projection operation.
As mentioned previously, the construct \mintinline{text}{tile.MN} is first converted to the call \mintinline{text}{Base.getproperty(tile, :MN)}.
The type of the second argument, \mintinline{text}{:MN}, is a \mintinline{text}{Symbol}, indicated by the colon prefix.
\mintinline{text}{Symbol}s are similar to strings, except that they are immutable and only one copy of each distinct value is stored~\cite{julia2020docs}.
The \mintinline{text}{Base.getproperty} function is specialised for arguments of type \mintinline{text}{Tile} on line 1.
The value of the \mintinline{text}{sym} argument of this function determines the name of the field that was accessed.
To generate custom projection implementations for each set of dimensions, we want to dispatch on the \emph{value} \mintinline{text}{:MN} of this argument, rather than its \emph{type} \mintinline{text}{Symbol}.
To do this, we can use Julia's \mintinline{text}{Val} type, a parametric type with one type parameter.
When we call the constructor of \mintinline{text}{Val} as \mintinline{text}{Val(sym)}, a new instance of \mintinline{text}{Val} is created where the type parameter is set to \mintinline{text}{sym}.
This essentially moves the value of \mintinline{text}{sym} to the type domain, so that we may use the multiple dispatch mechanism.
After creating a \mintinline{text}{Val} type, we dispatch to another function \mintinline{text}{getproperty_impl} that implements the projection itself.

\begin{listing}
\begin{minted}{julia}
@inline Base.getproperty(tile::Tile{size, names, T},​sym::Symbol) where {size, names, T} =​getproperty_impl(tile, Val(sym))

@generated function getproperty_impl(​tile::Tile{size, names, T}, ::Val{sym})​where {size, names, T, sym}
  if sym == :base || sym == :offset
    # standard fields
    return quote
      getfield(tile, sym)
    end
  else
    # tile projection
    sym_str = String(sym)
    new_names = ntuple(​i -> Symbol(sym_str[i]), length(sym_str))

    return quote
      # create new NamedTuples with the correct dimensions
      new_base   = ...
      new_offset = ...
      new_size   = ...

      # return projected tile
      return Tile{new_size, new_names, ...}(​new_base, new_offset)
    end
  end
end
\end{minted}
    \caption{Tile projection overview in our tiling API.}%
    \label{lst:tiling_project}
\end{listing}

To make the abstraction zero-cost, we use \mintinline{text}{@generated} functions that generate custom code at type-inference time and depending on the argument types, as shown on line 3 of \Cref{lst:tiling_project}.
Since we moved the field name to the type domain, we can thus generate a different, specialised implementation for each projection.
First note that accesses to the base or offset of a tile using \mintinline{text}{tile.base} or \mintinline{text}{tile.offset} also get converted to calls to \mintinline{text}{Base.getproperty}.
Lines 4--8 handle this by checking if the passed symbol is \mintinline{text}{base} or \mintinline{text}{offset}.
If so, we just return the value of the field by calling \mintinline{text}{getfield}.
Julia's \mintinline{text}{@generated} functions must return an \mintinline{text}{Expr}, which is a block of code to be compiled. Such blocks are surrounded with the \mintinline{text}{quote ... end} construct, as shown in lines 6--8.

The projection itself is implemented in lines 10--22.
Line 11 converts the symbol representing the field name to a \mintinline{text}{String}, which line 12 then converts to a tuple containing the individual dimensions.
For example, if \mintinline{text}{sym} is \mintinline{text}{:MN}, then \mintinline{text}{sym_str} and \mintinline{text}{new_names} are \mintinline{text}{"MN"} and \mintinline{text}{(:M, :N)}, respectively.
In lines 16--18, an \mintinline{text}{Expr} is generated to create new \mintinline{text}{NamedTuple}s that only contain the relevant dimensions for the base, offset, and size.
Finally, line 21 wraps these newly generated \mintinline{text}{NamedTuple}s in the \mintinline{text}{Tile} struct that represents the projected tile, and returns that tile.

The parallelise operation is exposed as a function call \mintinline{text}{parallelise(tile, tiling_size, index, count)}. The \mintinline{text}{tile} argument of type \mintinline{text}{Tile} is the parent tile that will be subdivided and parallelised over a set of entities that can be blocks, warps, or threads that cooperate.
The second argument, \mintinline{text}{tiling_size}, determines the tile size that each entity will handle, and the last argument \mintinline{text}{count} refers to the number of cooperating entities.
Finally, the argument \mintinline{text}{index} is an integer from 0 to \mintinline{text}{count - 1}, and determines the identifier of the currently executing entity.

\begin{figure}[t]
    \centering
  \includegraphics[width=1.8cm]{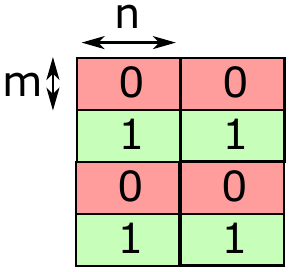}
    \caption{Parallelisation over 2 warps each handling a \( 4 \times 2 \) set of subtiles.}%
    \label{fig:tiling-parallellise-2}
\end{figure}

\Cref{fig:tiling-parallellise-2} shows an example parallelisation.
It starts with a parent tile of size \( 4m \times 2n \), divides it in subtiles of size \( m \times n \), and parallelises them across 2 warps.
The 0/1 in each subtile indicates the warp responsible for it.
We write the operation as \mintinline{text}{parallelise(Tile(M = 4 * m, N = n), Tile(M = m, N = n), warpId, 2)}, where \mintinline{text}{warpId} is either 0 or 1, i.e.\ the id of the currently executing warp.

To generalise the parallelisation operation to multiple dimensions, we again reuse the indexing functionality from Julia's standard library.
The information needed for iteration is then stored in a new struct, a \mintinline{text}{TileIterator}, that is returned by the \mintinline{text}{parallelise} function.
Julia allows us to write customised implementations for iterating over user-defined types.
For-loops are converted to calls to the \mintinline{text}{Base.iterate} function, which may be specialised for our own types.
To iterate over \mintinline{text}{TileIterator}s using a for loop, we must thus specialise the \mintinline{text}{Base.iterate} method for \mintinline{text}{TileIterator}s.
\mintinline{text}{Base.iterate} is called for each iteration of the for loop, and must return the value associated with each iteration.
In the case of \mintinline{text}{TileIterator}s, each call to \mintinline{text}{Base.iterate} will return a \mintinline{text}{Tile} corresponding to the tile of that iteration.

All operations on \mintinline{text}{Tile}s in our API are built on top of Julia interfaces that work for any number of dimensions.
For example, the position and size of each \mintinline{text}{Tile} is stored using Julia's \mintinline{text}{NamedTuple}s, which support any amount of dimensions.
Similarly, the parallelisation and linearisation operations, which involve computations using multidimensional indices, are written using Julia's generic indexing interfaces. This supports higher dimensions as required.

\subsection{Example Usage}
To illustrate the use, readability and zero cost of the API, we consider three representative stages of the tiled GEMM.

\subsubsection{Copying a tile of C from global to shared memory}

\begin{figure}[t]
    \centering
  \includegraphics[width=7cm]{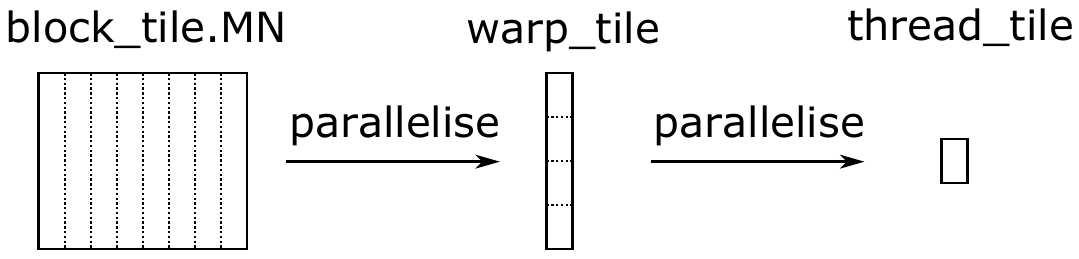}
    \caption{Copying a tile of the \( C \) matrix from global to shared memory.}%
    \label{fig:tiling-example-1}
\end{figure}

To copy a tile of \( C \) from global to shared memory in step 1 of the complete \gls{GEMM},  \Cref{lst:tiling_example_1} implements the approach illustrated in \Cref{fig:tiling-example-1}. Each block copies a separate tile, and we launch the \gls{GEMM} kernel with enough blocks to fully cover the \( C \) matrix.
The tile size is determined by the \mintinline{text}{block_tile} variable. It initially has three dimensions, so we first project it to the \( M \) and \( N \) dimension using \mintinline{text}{block_tile.MN} on line~1 in \Cref{lst:tiling_example_1}.

Next, we divide \mintinline{text}{block_tile} in subtiles and parallelise the resulting \mintinline{text}{warp_tile}s over a set of \mintinline{text}{WARPS_PER_BLOCK} cooperating warps in the block, also on line~1. The \mintinline{text}{@unroll} macro from the Julia package \term{GPUifyLoops.jl}~\cite{churavy2020gpuifyloops} informs LLVM to fully unroll the loop.
Each of these \mintinline{text}{warp_tile}s has size \mintinline{text}{(M = MEM_CD_WARP.M, N = MEM_CD_WARP.N)}.

Typically, \mintinline{text}{MEM_CD_WARP.N} is \( 1 \), so that the resulting \mintinline{text}{warp_tile} is highly rectangular.
This is necessary to access global memory efficiently, as this guarantees that the threads in one warp access adjacent memory locations.
The hardware is then able to coalesce these memory accesses into fewer memory transactions, thus increasing memory throughput.
This is commonly referred to as \emph{global memory coalescing}.

Similarly, we parallelise the \mintinline{text}{warp_tile} over the set of 32 threads in a warp on line~2.
The integer variable \mintinline{text}{laneId} identifies the threads within a warp.
Each thread handles a tile of size \mintinline{text}{(M = MEM_CD_THREAD.M, N = MEM_CD_THREAD.N)} in each iteration.
In the case of an \gls{FP32} \( C \) matrix, the best choice is \mintinline{text}{MEM_CD_THREAD.M = 4}, and \mintinline{text}{MEM_CD_THREAD.N = 1}.
This way, each thread loads/stores 4 adjacent \gls{FP32} elements, such that the GPU can issue one 128-bit load/store, the largest memory transaction size supported by the \gls{GPU}, thus maximally vectorising the memory accesses.

\begin{listing*}
\begin{minted}{julia}
@unroll for warp_tile = parallelise(block_tile.MN, Tile(MEM_CD_WARP), warpId, WARPS_PER_BLOCK)
  @unroll for thread_tile = parallelise(warp_tile, Tile(MEM_CD_THREAD), laneId, 32)
    global_thread_tile = translate(thread_tile, (M = block_i, N = block_j))

    global_linear_base   = linearise(global_thread_tile.base, (M = global_M, N = global_N))
    global_linear_offset = linearise(global_thread_tile.offset, (M = global_M, N = global_N))
    shared_linear_base   = linearise(thread_tile.base, (M = shared_M, N = shared_N))
    shared_linear_offset = linearise(thread_tile.offset, (M = shared_M, N = shared_N))

    global_ptr = pointer(global_c, global_linear_base)
    shared_ptr = pointer(shared_c, shared_linear_base)

    value = vloada(Vec{MEM_CD_THREAD, Float32}, global_ptr, global_linear_offset)
    vstorea!(Vec{MEM_CD_THREAD, Float32}, shared_ptr, value, shared_linear_offset)
  end
end
\end{minted}
    \caption{Copying a tile of the \( C \) matrix from global to shared memory using our tiling API.}
    \label{lst:tiling_example_1}
\end{listing*}

The positions of all tiles are specified relative to the top-left corner of the current block's tile.
This means that \mintinline{text}{thread_tile.index == (M = 0, N = 0)} corresponds to a linear index of 0.
Because shared memory only stores the tile of the current block, this is the correct index for shared memory.
For global memory, we need to offset this tile depending on the currently executing block.
To accomplish this, we translate this \mintinline{text}{thread_tile} over the correct distance on line~3.
Finally, lines~5--8 convert the base and offset of each of these \mintinline{text}{thread_tile}s to a linear index.
We can then create a pointer to the correct memory location on lines~10--11, and perform the load or store.
To separate the constant parts of the memory addresses, we create a pointer using the linearised base, and only add the linearised offset afterwards.

\Cref{lst:tiling_manual} is equivalent to \Cref{lst:tiling_example_1}, but does not use our tiling API.
The \mintinline{text}{BLOCK_M} and \mintinline{text}{BLOCK_N} variables on line~1 are constants that correspond to the size of \mintinline{text}{block_tile} in \Cref{lst:tiling_example_1}.
The outer loop on lines 1--6 corresponds to the first parallelisation, the inner loop on lines 8--13 is the equivalent of the second parallelisation.
In both loops the tile bases and offsets are calculated manually.
Lines 15--18 convert the bases and offsets to linear indices, and are thus the equivalent of the linearisations in \Cref{lst:tiling_example_1}. The translation is handled by the addition of the translation offsets \texttt{block\_i} and \texttt{block\_j} on line~15.
Clearly the use of our tiling \gls{API} in \Cref{lst:tiling_example_1} is less verbose and more maintainable.

\begin{listing*}
\begin{minted}{julia}
@unroll for warp_offset = 0 : WARPS_PER_BLOCK : (BLOCK_M * BLOCK_N) ÷ (MEM_CD_WARP.M * MEM_CD_WARP.N) - 1
  NUM_WARP_ROWS = BLOCK_M ÷ MEM_CD_WARP.M
  base_warp_i = (warpId % NUM_WARP_ROWS) * MEM_CD_WARP.M
  base_warp_j = (warpId ÷ NUM_WARP_ROWS) * MEM_CD_WARP.N
  warp_i = (warp_offset % NUM_WARP_ROWS) * MEM_CD_WARP.M
  warp_j = (warp_offset ÷ NUM_WARP_ROWS) * MEM_CD_WARP.N

  @unroll for thread_offset = 0 : 32 : (MEM_CD_WARP.M * MEM_CD_WARP.N) ÷ (MEM_CD_THREAD.M * MEM_CD_THREAD.N) - 1
    NUM_THREAD_ROWS = MEM_CD_WARP.M ÷ MEM_CD_THREAD.M
    base_thread_i = (laneId % NUM_THREAD_ROWS) * MEM_CD_THREAD.M
    base_thread_j = (laneId ÷ NUM_THREAD_ROWS) * MEM_CD_THREAD.N
    thread_i = (thread_offset % NUM_THREAD_ROWS) * MEM_CD_THREAD.M
    thread_j = (thread_offset ÷ NUM_THREAD_ROWS) * MEM_CD_THREAD.N

    global_linear_base   = (block_i + base_warp_j + base_thread_j) * global_M + (block_j + base_warp_i + base_thread_i)
    global_linear_offset = (warp_j + thread_j) * global_M + (warp_i + thread_i)
    shared_linear_base   = (base_warp_j + base_thread_j) * shared_M + (base_warp_i + base_thread_i)
    shared_linear_offset = (warp_j + thread_j) * shared_M + (warp_i + thread_i)

    global_ptr = pointer(global_c, global_linear_base)
    shared_ptr = pointer(shared_c, shared_linear_base)

    value = vloada(Vec{MEM_CD_THREAD, Float32}, global_ptr, global_linear_offset)
    vstorea!(Vec{MEM_CD_THREAD, Float32}, shared_ptr, value, shared_linear_offset)
  end
end
\end{minted}
    \caption{Implementing the first stage in GEMM using manual calculation of addresses.}
    \label{lst:tiling_manual}
\end{listing*}

\Cref{lst:tiling_example_1_ptx} shows the CUDA \gls{PTX} to which \Cref{lst:tiling_example_1} is compiled.
First, each thread's base addresses are computed in registers \texttt{\%rd20} and \texttt{\%rd13} for shared and global memory, respectively.
The loads and stores are vectorised, as indicated by the suffix \texttt{v4.f32}.
The stores to shared memory are on lines~6, 12, and 20. As the shared memory size is known at compile time, the code exploits the register plus constant addressing modes as discussed in~\Cref{sub:tiling_API}.
By contrast, the compiler does not know the size of the matrix in global memory, so it does not know the linearised offset either, even though the offsets in the \( M \) and \( N \) dimensions are constants.
To calculate the address in global memory, LLVM emits a multiplication (using a bit shift \texttt{shl.b64}), and an addition.

The code in \Cref{lst:tiling_example_1_ptx} is identical to the code that the Julia-LLVM tool flow generates for \Cref{lst:tiling_manual}.
We conclude that no superfluous instructions are generated because of the use of our tiling API, for both the loads from global memory and the stores to shared memory.

We can use similar code for steps 2, 3-a, 3-b, 3-c-i, and 3-c-ii of the \gls{GEMM}, with independently chosen tile configurations for each of them. This way, recursive double-sided blocking is supported.

\begin{listing}
\begin{minted}{ptx.py:CustomLexer -x}
// Calculate the base addresses in %rd13 and %rd20...
shl.b64            %rd22, %rd13, 5;
add.s64            %rd23, %rd17, %rd22;
cvta.to.global.u64 %rd24, %rd23;
ld.global.v4.f32   {%f5, %f6, %f7, %f8}, [%rd24];
st.shared.v4.f32   [%rd20+4096], {%f5, %f6, %f7, %f8};

shl.b64            %rd25, %rd13, 6;
add.s64            %rd26, %rd17, %rd25;
cvta.to.global.u64 %rd27, %rd26;
ld.global.v4.f32   {%f9, %f10, %f11, %f12}, [%rd27];
st.shared.v4.f32   [%rd20+8192], {%f9, %f10, %f11, %f12};

// ... repetition of similar blocks due to unrolling

mul.lo.s64         %rd64, %rd13, 480;
add.s64            %rd65, %rd17, %rd64;
cvta.to.global.u64 %rd66, %rd65;
ld.global.v4.f32   {%f61, %f62, %f63, %f64}, [%rd66];
st.shared.v4.f32   [%rd20+61440], {%f61, %f62, %f63, %f64};
\end{minted}
    \caption{The PTX code generated for \Cref{lst:tiling_example_1}.}
    \label{lst:tiling_example_1_ptx}
\end{listing}

\begin{figure}[t]
    \centering
  \includegraphics[width=5.4cm]{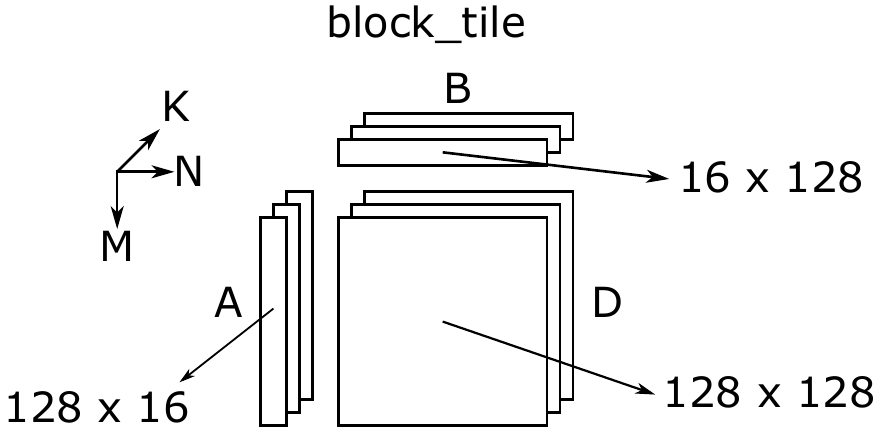}
    \caption{3D iteration space in the inner loop of the matrix product.}%
    \label{fig:tiling-example-2-block}
\end{figure}

\subsubsection{Computation of the matrix product}

\begin{figure}[t]
    \centering
  \includegraphics[width=6.3cm]{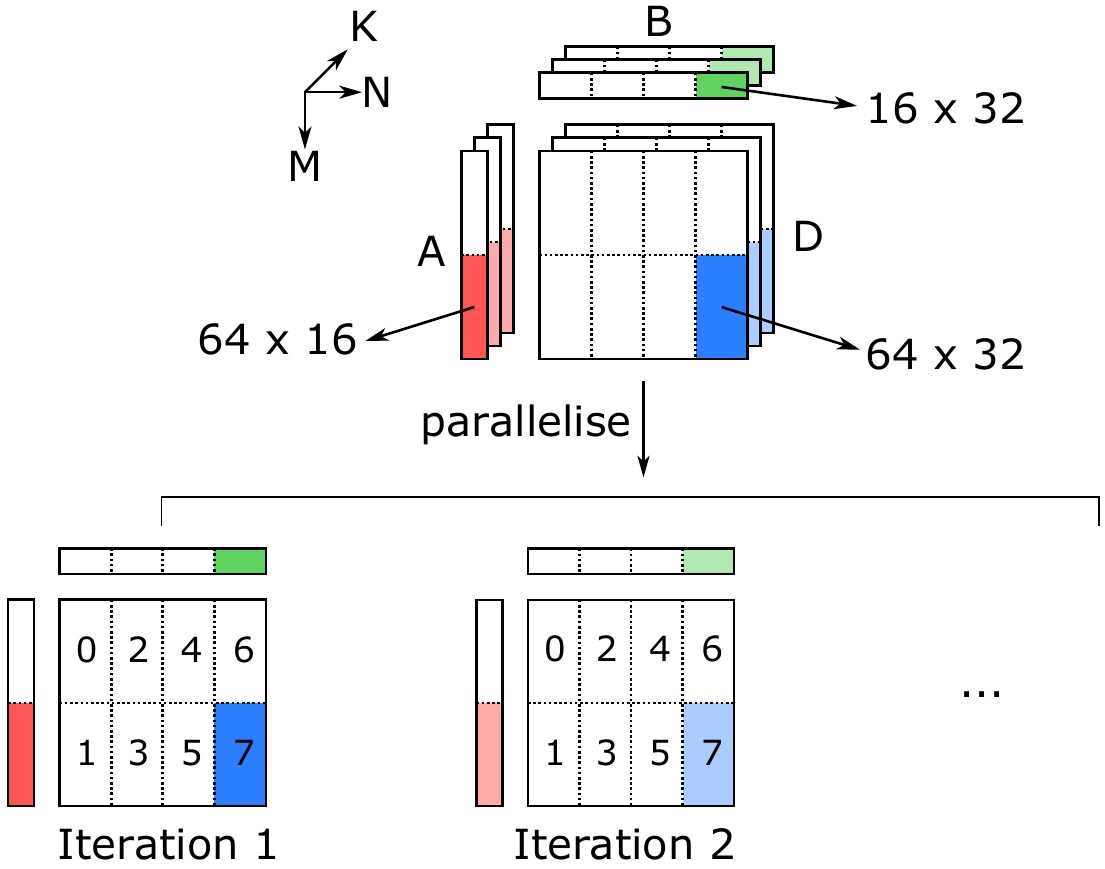}
    \caption{Computation of the matrix product in the innermost loop.}%
    \label{fig:tiling-example-2}
\end{figure}

To implement the computation of the matrix product in the inner loop using the tiling \gls{API}, we will follow the approach illustrated in \Cref{fig:tiling-example-2-block}.
A \mintinline{text}{block_tile} represents the three-dimensional iteration space \( (M, N, K) \) used to calculate the tile of the \( D \) matrix corresponding to one block.
Let us consider the case where a \mintinline{text}{block_tile} has size \( (M,N,K)=(128,128,16) \).
This means that each block calculates an \( M \times N = 128 \times 128 \) tile of \( D \), by multiplying all \( M \times K = 128 \times 16 \) tiles in a row of \( A \) with all \( K \times N = 16 \times 128 \) tiles in a column of \( B \).
These tiles of \( D \) are subsequently accumulated by summing over the \( K \) dimension.

We want to parallelise this computation over all warps in a block.
In the example of \Cref{fig:tiling-example-2}, each block contains 8 warps, in a \( 2 \times 4 \) arrangement.
Each warp calculates a \( 64 \times 32 \) tile of \( D \) in each iteration, by multiplying a \( 64 \times 16 \) tile of \( A \), and a \( 16 \times 32 \) tile of \( B \).
Of course, we want tiles in this three-dimensional space with the same \( M \) and \( N \) indices to be mapped to the same warp, so that we can accumulate across the \( K \) dimension.
In the case where the matrices are stored in column-major, the warps are assigned to tiles in the order of the \( M \), \( N \), and \( K \) dimension.
We can thus simply use a parallelisation operation of size \( (M, N, K) = (64, 32, 16) \) across 8 warps, as shown in \Cref{fig:tiling-example-2}.
The 8 warps fully cover the \( M \) and \( N \) dimensions, as indicated by the 0--7 in each tile.
In the next iteration, we have advanced along the \( K \) dimension, but the division along the \( M \) and \( N \) dimension is the same, so with this choice of tiling size, the parallelisation operation implicitly iterates over the \( K \) dimension.

Line 1 of \Cref{lst:tiling_example_2} shows this parallelisation operation. It returns a three-dimensional \mintinline{text}{warp_tile}.
To compute the matrix product using \gls{WMMA}, we first need to load the \( A \) and \( B \) tiles into \gls{WMMA} fragments.
To load \( A \), we are only interested in the \( M \) and \( K \) dimension, so we first project \mintinline{text}{warp_tile} on line~3.
This gives a tile of size \( (M, K) = (64, 16) \), which thus consists of four \( 16 \times 16 \) \gls{WMMA} fragments.
To load those, we first translate the tile in the \( M \) dimension over 0, 16, 32, and 48 elements on line~3.
Lines 5--6 then convert this translated base and offset to a linear index, which can then be used to create the pointer argument to \mintinline{text}{WMMA.load_a} on line~8.
Lines 11--18 do the same thing for the \( B \) matrix: the \mintinline{text}{warp_tile} is projected to the \( K \) and \( N \) dimensions, translated, and converted to a linear index.
Finally, lines 20--24 calculate the \( 64 \times 32 \) product of \( D \) using the \mintinline{text}{WMMA.mma} function from our WMMA API.

\begin{listing}
\begin{minted}{julia}
@unroll for warp_tile = parallelise(block_tile,​Tile(M=64, N=32, K=16), warpId, 8)
  @unroll for i = 1 : 4
    a_tile = translate(warp_tile.MK, (M=(i-1)*16, K=0))
    linear_base = linearise(a_tile.base, ...)
    linear_offset = linearise(a_tile.offset, ...)
    a_frags[i] = WMMA.load_a(...)
  end
  @unroll for j = 1 : 2
    b_tile = translate(warp_tile.KN, (K=0, N=(j-1)*16))
    linear_base = linearise(b_tile.base, ...)
    linear_offset = linearise(b_tile.offset, ...)
    b_frags[j] = WMMA.load_b(...)
  end
  @unroll for i = 1 : 4
    @unroll for j = 1 : 2
      acc_frags[i, j] = WMMA.mma(a_frags[i], b_frags[j],​acc_frags[i, j], ...)
    end
  end
end
\end{minted}
    \caption{Matrix product computation using our tiling API.}
    \label{lst:tiling_example_2}
\end{listing}

This example is perhaps the best illustration of the tiling \gls{API}, as it combines all four operations on tiles: parallelisation, projection, translation, and linearisation.
Using these four operations significantly improves readability compared to writing the necessary address calculations by hand.

We omit the \gls{PTX} code generated for this listing because it provides no additional value, but we confirm similar observations as on the first example: the base addresses of \( A \) and \( B \) for each warp are calculated once, and stored in registers.
The code in \Cref{lst:tiling_example_2} is converted to a set of \texttt{wmma.load.a}, \texttt{wmma.load.b}, and \texttt{wmma.mma} instructions, and the addresses of the load operations are expressed as a constant offset from the base addresses stored in registers.
This once again indicates that the tiling abstractions do not introduce any superfluous instructions.

\subsubsection{Copying a tile of D from registers to shared memory}

\begin{figure}[t]
    \centering
  \includegraphics[width=7.5cm]{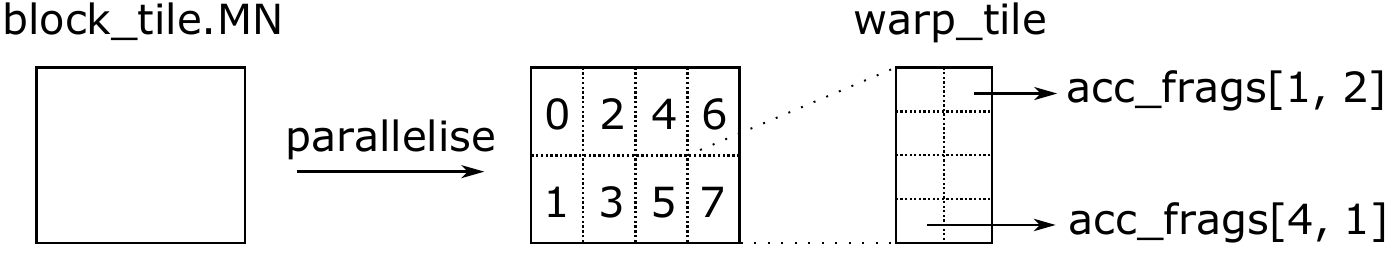}
    \caption{Copying a tile of \( D \) from registers to shared memory.}%
    \label{fig:tiling-example-3}
\end{figure}

In the previous example, we studied the calculation of the matrix product in the inner loop of \gls{GEMM}.
After this stage, each warp has a part of the \( D \) matrix stored in \gls{WMMA} fragments.
To store these \gls{WMMA} fragments to shared memory, we follow the approach illustrated in \Cref{fig:tiling-example-3}.
\texttt{block\_tile} represents the same tile as in example 2, i.e.\ the three-dimensional iteration space used to calculate a tile of the \( D \) matrix corresponding to one block.
To copy \( D \), we are only interested in the \( M \) and \( N \) dimension, so we project this tile to these dimensions first.

Next, we parallelise this tile over a set of warps.
This parallelisation should have the same parameters as the matrix computation in the previous example.
Obviously, the tiling size is only specified in the \( M \) and \( N \) dimension, instead of in the three dimensions \( M \), \( N \), and \( K \).
\Cref{fig:tiling-example-3} uses the same tiling sizes as our previous example: \texttt{block\_tile} is a \( 128 \times 128 \) matrix, and is parallelised across \mintinline{text}{WARPS_PER_BLOCK} = 8 warps, each handling a \mintinline{text}{COMPUTE_WARP.M} \( \times \) \mintinline{text}{COMPUTE_WARP.N} = \( 64 \times 32 \) subtile.
The corresponding parallelisation operation returns a \texttt{warp\_tile}, and is shown on line~1 of \Cref{lst:tiling_example_3}.
Note that the for loop of line~1 only has 1 iteration in this case, since 8 warps fully cover the \texttt{block\_tile}.

Finally, this \texttt{warp\_tile} is divided in a \( 4 \times 2 \) arrangement of \gls{WMMA} fragments, like in example 2.
The for loops on line~2 and line~3 iterate over these 8 \gls{WMMA} fragments.
Line~4 then translates the tile in the \( M \) and \( N \) dimension over \( 0 \), \( 16 \), \( 32 \), or \( 48 \) elements to obtain the final tile corresponding to each \gls{WMMA} fragment.
Line~6 and Line~7 then convert this Cartesian index to a linear index, so that it may be used to create pointers for the \gls{WMMA} \texttt{store.d} on line~9.

Again, we omit the generated \gls{PTX} code, but we can confirm our earlier observations. First, the base address of \( D \) for each warp is calculated and stored in a register. After this computation, 8 \texttt{wmma.store.d} instructions are emitted, which use this register as a base address, and constant offsets. Once again, we conclude that the use of the tiling \gls{API} does not introduce any extra overhead.

\begin{listing}
\begin{minted}{julia}
@unroll for warp_tile = parallelise(block_tile.MN, Tile(COMPUTE_WARP).MN, warpId, WARPS_PER_BLOCK)
  @unroll for i = 1 : 4
    @unroll for j = 1 : 2
      tile = translate(warp_tile, (M=(i-1)*16, N=(j-1)*16))

      linear_base   = linearise(tile.base, ...)
      linear_offset = linearise(tile.offset, ...)

      WMMA.store_d(..., acc_frags[i, j], ...)
    end
  end
end
\end{minted}
    \caption{Copying a \( D \) tile from registers to shared memory.}
    \label{lst:tiling_example_3}
\end{listing}




\section{Flexible GEMM kernel abstractions}%
\label{sec:matmul}
In the previous section, we designed tiling abstractions to
implement performant GEMM kernels. In this section, we add the
necessary flexibility to this GEMM, such that users can instantiate a wide range
of GEMM variants.

\subsection{Requirements}%
\label{sub:GEMM_requirements}

The Google Brain \gls{DL} research team provides an excellent overview of why this flexibility is needed~\cite{barham2019machine}. They focus on Capsule networks, a novel neural network \gls{ML} idea where the neurons are matrix-valued rather than scalars~\cite{hinston2018matrix}. In short, they observed that the inflexibility of existing \gls{ML} frameworks TensorFlow~\cite{abadi2016tensorflow} and PyTorch~\cite{paszke2019pytorch} forced the researches to rephrase their computations in terms of the limited set of GEMM kernels already supported by these frameworks. They had to insert multiple data transposition and matrix materialisation stages that introduced detrimental amounts of memory access overhead. They also had to insert separate kernels in between their network layers to perform simple but not-yet-established operations on the matrix elements. Not being able to fuse those operations in the GEMM kernels themselves, this again introduced massive amounts of overhead. Clearly, for advanced research in a domain such as ML, libraries providing only established GEMM functionality do not suffice.

A flexible \gls{GEMM} also needs to support a multitude of different memory layouts.
Basic \gls{GEMM}s involve only row-major and column-major layouts. Convolutions, which are also implemented with GEMMs, involve more dimensions than matrices, so more layouts need to be considered. For images, e.g.\ \gls{ML} frameworks typically use four dimensions: a batch of \( N \) images with \( C \) channels, each consisting of \( W \times H \) features. Among the many possible choices, NCHW and NHWC are most common~\cite{nvidia2020deep}.

Next, we consider the generalisation of matrix multiplications to multidimensional \glspl{TCT}, which are common in several scientific fields, such as fluid dynamics~\cite{rink2018cfdlang}, electromechanics~\cite{poya2017ahighperformance}, and computational chemistry~\cite{auer2006automatic}. Whereas a matrix-matrix multiplication has three different indices \( \{ m, n, k \} \), \glspl{TCT} involve an arbitrarily large set of indices. Matrix transpositions are extended to arbitrary permutations of those indices.
The number of possible data layouts for \glspl{TCT} is hence much higher. For example, a \gls{TCT} of 4D tensors has a total of \( 4! \times 4! \times 4! = 13824 \) different memory layouts.

Given the importance of \glspl{TCT}, a lot of research has been done to implement efficient support for the large number of possible cases.
Springer and Bientinesi classify the traditional approaches to \glspl{TCT} in three main categories~\cite{springer2017landscape}: loop nesting~\cite{apra2014efficient,ma2011gpu,nelson2015generating,springer2016design}, \acrshort{LoG}~\cite{napoli2014towards,li2015aninput}, and \gls{TTGT}~\cite{solomonik2013cyclops,bader2006algorithm}. All three of them suffer from serious performance issues due to bad data reuse or the need to insert data reshuffling and transposition kernels.

In 2016, Springer and Bientinesi proposed another method for \glspl{TCT}, \gls{GETT}~\cite{springer2016design}, that has since been adopted by other \gls{TCT} implementations~\cite{kim2019acode,matthews2018high}.
\gls{GETT} is based on the principles of \gls{TTGT}, but implicitly reorganises tensors while loading them to avoid separate transpositions.
\gls{GETT} can therefore be seen as a variant of \gls{TTGT}, where the transposes are fused into the \gls{GEMM} kernel.
Clearly, this fusion requires that the underlying \gls{GEMM} kernel is flexible.

While the tiling and WMMA APIs introduced in previous sections allow that flexibility, programming against them would still be quite cumbersome. We hence propose a higher-level API based on a high-performance GEMM kernel that can easily be customised through a set of higher-level abstractions that are as intuitive as possible to researchers from, e.g.\ the domains of \gls{ML} and \gls{DL}. We put forward the following main requirements for this high-level GEMM API:
\begin{itemize}
\item \emph{Flexibility} with respect to data layouts, on-the-fly data transpositions, and fused operations are prime objectives. Support for other, more complex data types such as complex numbers~\cite{abdelfattah2019towards} or dual numbers as used in automatic differentiation~\cite{revels2016autodiff} is also of interest.
\item \emph{Performance} of \gls{GEMM} kernels that are built using our \gls{API} should be on-par with the state-of-the-art implementations, such as \term{cuBLAS} or \term{CUTLASS}. This obviously implies that our GEMM should be WMMA-compatible and support double-sided recursive blocking, i.e.\ independent tiling parameters need to be supported for the data transfer stages at different levels of the memory hierarchy and for the computational stages.
\item \emph{Portability}: \gls{GEMM} kernels built with our \gls{API} should perform well on a range of devices.
        We should hence make as few assumptions about the underlying hardware as possible. For example, our \gls{API} needs to be able to handle different shared memory sizes, as well as GPUs with and without \glspl{TC} of different generations.
\end{itemize}

\subsection{A flexible, abstract GEMM API for Julia}%
\label{sub:abstract_components}

Our strategy is to implement the general structure of a performant \gls{GEMM} kernel beforehand.
To make it flexible, we split this \gls{GEMM} in a small set of building blocks with a predetermined interface. Concretely, the GEMM contains calls to a set of functions with a predetermined name. To extend the basic implementation, it will suffice to implement new versions of the called functions that customise the behaviour based on their input types. Julia's just-in-time type inference and compilation flow enables us to perform this split without introducing performance overhead. Furthermore, Julia's multiple dispatch allows us to make the split orthogonally, which results in more intuitive building blocks and eases code reuse and hence programmer productivity.

\subsubsection{Params}%
\label{ssub:params}
We of course still want the user to be able to customise the tiling size of each step of the \gls{GEMM} kernel.
This is the purpose of the \emph{params} abstraction.
This abstraction is essentially a structure that is passed to the kernel, and contains a set of configuration fields.
Some of these fields determine the tiling sizes, others specify the kernel's launch configuration, such as the number of warps per block.
The user does not need to specify all fields manually.
We have implemented a set of heuristics that choose reasonable defaults for fields that are not set explicitly.
For example, if the tiling size per threadblock is not set, we choose the largest square (\( N \times N \)) or nearly-square (\( 2N \times N \)) tile that still fits in shared memory.
For the time being, these heuristics are mainly aimed at \glspl{GEMM} using \glspl{TC}, but future work could expand these heuristics to other cases as well.

\subsubsection{Layouts}%
\label{ssub:layouts}
The positions of tiles at different levels in the memory hierarchy in our tiling \gls{API} are expressed in logical coordinates.
To convert these logical coordinates to offsets in physical memory, we introduce another abstraction, called \emph{layouts}.
This abstraction corresponds to three functions that can be customised using Julia's multiple dispatch.
The \mintinline{text}{size(layout_type, logical_size)} function determines the size in physical memory of the layout for a given size in logical coordinates.
This physical size is not necessarily the same as the size in logical coordinates.
For example, to access shared memory efficiently, it is sometimes necessary to add \( p \) padding elements to every column of a column major matrix.
In this case, for a logical size of \( M \times K \), the corresponding physical size would be \( (M + p) \times K \).
The \mintinline{text}{size(...)} function is used so that our \gls{GEMM} \gls{API} knows how many bytes it has to reserve in shared memory.
This function is also used by the heuristics in the params abstraction to select the optimal tiling size in shared memory, as this depends on how much memory a given memory layout requires.

The other two functions are \mintinline{text}{load(layout_type, tile, ...)} and \mintinline{text}{store(layout_type, tile, ...)}.
As their name suggests, these functions are responsible to load or store the tile at the logical coordinates represented by the \mintinline{text}{tile} argument.
With these functions, users can implement arbitrary logic to load or store the matrix elements corresponding to a given tile.
For example, recall that NVIDIA \glspl{GPU} can load and store vectors of 16 bytes (128 bits) in a single instruction.
This vectorisation of memory accesses is only possible if the base address of the load or store is aligned, i.e.\ divisible by 16.
An \mintinline{text}{AlignedColumnMajor} layout can indicate that the necessary alignment requirements are met, so that the corresponding \mintinline{text}{load} and \mintinline{text}{store} functions can issue vectorised loads and stores.

For a classic \gls{GEMM} kernel, the most obvious instantiations of the layout building block are \mintinline{text}{RowMajor} and \mintinline{text}{ColumnMajor}.
As mentioned before, each of these can be adapted to aligned or padded layouts.
To add padding, one could have \mintinline{text}{PaddedRowMajor} and \mintinline{text}{PaddedColumnMajor} layouts, but Julia's type system allows us to do this more cleanly.
We can make a parameterised type \mintinline{text}{PaddedLayout{Layout, Padding}}, where \mintinline{text}{Padding} represents the padding in number of elements, and \mintinline{text}{Layout} is the base layout we wish to modify, such as \mintinline{text}{RowMajor} or \mintinline{text}{ColumnMajor}.
The \mintinline{text}{load} and \mintinline{text}{store} functions for padded layouts would then dispatch to the implementations for the underlying \mintinline{text}{Layout}.

The layout building block can also be used to create a \gls{GEMM} with a more complicated mapping between logical indices and physical offsets.
For example, \gls{GETT}'s reinterpretation of multidimensional tensors as matrices can be performed using a custom implementation of the layout building block.
Note that a layout does not even need to correspond to a matrix that is materialised in memory.
Consider a matrix multiplication where the elements of one of the matrices can be calculated from the position, i.e.\ \( a_{ij} = f(i, j) \) for some function \( f \).
In this case, we implement a layout where the \mintinline{text}{store} function is a no-op, and the \mintinline{text}{load} function generates the necessary elements on the fly.
Similar strategies can be used for other matrices with a special structure, such as sparse matrices or diagonal matrices.
We can only store the non-zero elements in memory, and create a custom layout that implements the necessary logic to load or store the correct elements.

\begin{figure}[t]
    \centering
    \includegraphics[width=6.7cm]{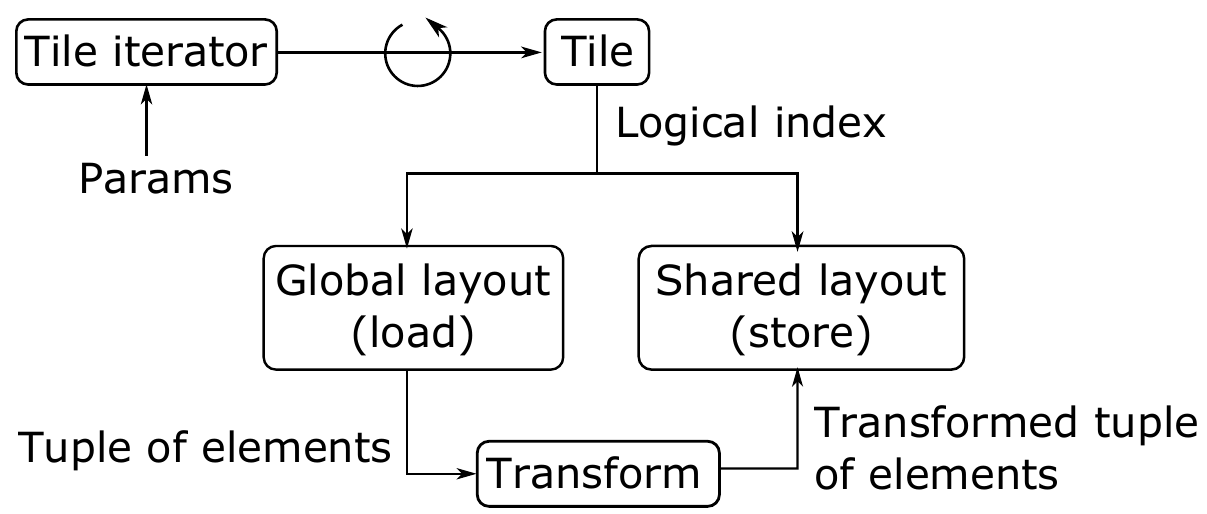}
    \caption{Copying an \( A \) tile from global to shared memory.}%
    \label{fig:gemm-abstractions}
\end{figure}

\subsubsection{Transforms}%
\label{ssub:transforms}
The next building block is that of \emph{transforms}.
Transforms are arbitrary Julia functors, i.e.\ functions or structures implementing the function call operator \mintinline{text}{()}.
They are called after every load and before every store operation in the GEMM. By having a transform after every load and before every store, element-wise operations to the input and result matrices can be applied consistently in our \gls{API}.

Transforms can serve for element-wise operations, such as a simple scaling in the case of \gls{GEMM}, and for activation functions for artificial neurons in neural networks.
Another use case of transforms is to implement type conversions immediately after loading data from global memory.
This is useful if one wants to use a higher precision data type to compute the \gls{GEMM}, but store the matrices in lower precision in global memory to save capacity.

\cref{fig:gemm-abstractions} illustrates how params, layouts, and transforms interact to copy a tile of the \( A \) matrix from global to shared memory.
A similar structure is used to copy tiles of the \( B \), \( C \), or resultant \( D \) matrix.
This copy operation is performed cooperatively by all threads in a threadblock, using the parallelisation operation of our tiling \gls{API}.
First, the params component determines the tiling size that should be used for the tile iterator corresponding to the parallelise operation.
The \gls{GEMM} kernel then iterates over this tile iterator, which returns a tile in each iteration.
The base and offset of this tile are specified in logical coordinates.
To load the correct matrix elements from global memory, the \mintinline{text}{load} function is called using this tile and the layout of \( A \) in global memory.
This \mintinline{text}{load} function returns a tuple that contains the correct matrix elements.
This tuple is then sent to the transform for the global-to-shared stream of the \( A \) matrix, resulting in a transformed tuple.
Finally, the \mintinline{text}{store} function corresponding to the layout of \( A \) in shared memory is called with this transformed tuple and the logical index of the current tile.

\subsubsection{Operators}%
\label{ssub:operators}
The previous building blocks together copy tiles from global to shared memory.
The purpose of the next building block, called \emph{operators}, is to load tiles from shared memory, perform the matrix multiplication, and store the resulting tile back to shared memory. To do so, this building block has five functions associated with it.

The \mintinline{text}{load_a}, \mintinline{text}{load_b}, and \mintinline{text}{load_c} functions load tiles of the \( A \), \( B \), and \( C \) matrix from shared memory to registers.
The matrix computation itself is performed by the \mintinline{text}{mma} function, and the result is stored back to shared memory using the \mintinline{text}{store_d} function.
Like the layout building block, the \mintinline{text}{load_a}, \mintinline{text}{load_b}, \mintinline{text}{load_c}, and \mintinline{text}{store_d} functions have a \mintinline{text}{tile} argument that represents the logical coordinate of the tile that should be loaded or stored.
The load and store functions also have an argument that determines the shared memory layout of the corresponding matrix, so that we can dispatch to the different implementations depending on the memory layout that is used.
Finally, the \mintinline{text}{mma} function has three arguments \mintinline{text}{a_frag}, \mintinline{text}{b_frag}, and \mintinline{text}{c_frag} that represent parts of the \( A \), \( B \), and \( C \) matrices stored in registers.
The function should perform the multiply-accumulate operation \mintinline{text}{res_frag = a_frag * b_frag + c_frag}, and return the resulting fragment \mintinline{text}{res_frag}.

The listed functions map one-to-one onto the steps of the \gls{WMMA} API mentioned in~\Cref{sub:tensor_cores}.
This is no coincidence, as both the operator building block and the \gls{WMMA} \gls{API} are warp-level matrix multiply-accumulate operations.
It is hence fairly easy to define an implementation of the operator building blocks that uses \glspl{TC} using our \gls{WMMA} \gls{API}.
It suffices to convert the \mintinline{text}{tile} argument to the load and store functions to a memory address, and call the \mintinline{text}{load}, \mintinline{text}{store}, and \mintinline{text}{mma} functions of the \gls{WMMA} \gls{API}.

The operator building block has several use cases.
First, it can be used to provide a custom implementation for the computation in the inner loop of \gls{GEMM}.
This is useful if the data type of our matrices has a custom multiplication operator, such as complex numbers or dual numbers.
The operator building block also improves the portability of the \gls{GEMM} kernel. For example, the \gls{WMMA} operator may be parameterised with the \gls{WMMA} shape, so that we can select the \gls{WMMA} shape that is optimal for our \gls{GPU}. Alternatively, we can define an alternative operator that calculates the matrix product using the traditional \glspl{FPU} instead of \glspl{TC} on devices that lack the latter.

\subsubsection{Epilogues}%
\label{ssub:epilogues}
While the already discussed transform abstraction already allows performing element-wise operations on the \( D \) matrices, we add an epilogue abstraction to our API to enable customisation of the way global memory is updated at the last stage of \gls{GEMM}. This enables, e.g.\ to apply a reduction operation across all threadblocks.

In the general GEMM implementation, our epilogue building block only has one purpose: to copy tiles of the resultant matrix from shared memory to global memory.
By default, we only include one epilogue that simply copies the current threadblock's tile in shared memory to the correct position in global memory.
This default epilogue uses the previously mentioned layout building block to determine the memory layout of the resultant \( D \) matrix.

\subsection{Example Uses}%
\label{sub:usage}

\subsubsection{Fully-featured interface}%
\label{ssub:fully_featured_interface}
As a first example of how to use the presented building blocks and API, we consider the first step in a \gls{GEMM} kernel: copying a tile of the \( C \) matrix from global to shared memory.
\Cref{lst:tiling_example_1} showed an implementation of this step using our tiling \gls{API}. In our \gls{GEMM} \gls{API}, this first step is implemented as shown in \cref{lst:matmul_firststep}.
The code has a similar structure to \cref{lst:tiling_example_1}, but the linearisation, loads, and stores are replaced by generic calls to \mintinline{text}{Layout.load} and \mintinline{text}{Layout.store}.
The first arguments of these functions, \mintinline{text}{GLOBAL_LAYOUT} and \mintinline{text}{SHARED_LAYOUT}, are types that determine the memory layout of \( C \) for global and shared memory, respectively.
The \mintinline{text}{transform_global_to_shared_c} is a Julia function that represents the transform that should be applied during the global-to-shared memory stream of the \( C \) matrix.

\begin{listing}
\begin{minted}{julia}
@unroll for warp_tile = parallelise(block_tile.MN,​Tile(MEM_CD_WARP), warpId, WARPS_PER_BLOCK)
  @unroll for thread_tile = parallelise(warp_tile,​Tile(MEM_CD_THREAD), laneId, 32)
    global_thread_tile = translate(thread_tile,​(M=block_i, N=block_j))

    x = Layout.load(GLOBAL_C_LAYOUT, c, global_thread_tile)
    y = transform_global_to_shared_c(x, thread_tile)
    Layout.store(SHARED_C_LAYOUT, shmem_c, y, thread_tile)
  end
end
\end{minted}
    \caption{Copying a  \( C \) tile from global to shared memory.}
    \label{lst:matmul_firststep}
\end{listing}

Now suppose that we have defined the necessary components (such as layouts, operators, \ldots) for a given use case.
To instantiate and execute \gls{GEMM} kernels that use these components, we use the user-facing interface of our \gls{GEMM} \gls{API}, which is illustrated in \cref{lst:matmul_relu}.
This code fragment calculates a mixed-precision matrix product of the form \( D_{ij} = \max(\sum_k A_{ik} B_{kj} + C_{ij}, 0) \).
These types of matrix products are common in neural networks, where the activation function \( \max( \cdot, 0 ) \) is commonly referred to as a rectified linear unit (ReLU).
Lines 1--4 declare the two-dimensional arrays that represent the \( A \), \( B \), \( C \), and \( D \) matrices.
In lines 6--11, we configure the parameters of our \gls{GEMM} kernel, such as the overall shape of the \gls{GEMM}, the operator to be used in the inner loop, and the memory layouts of the \( A \) and \( C \) matrices.
The missing fields are automatically set to reasonable default values.
For example, if the memory layout of the \( B \) matrix is not specified, it is automatically set to the memory layout of the \( A \) matrix.

A \gls{GEMM} kernel using this configuration is executed in lines 13--16.
The transform that should be applied when copying tiles of the resultant \( D \) matrix from the register file to shared memory is determined by the argument \texttt{transform\_regs\_to\_shared\_d}.
The call to \mintinline{text}{GemmKernels.matmul} will execute each step in the \gls{GEMM} kernel, using the components given by the user.
For example, \cref{lst:matmul_firststep} will be executed with \mintinline{text}{GLOBAL_C_LAYOUT = Layout.AlignedColMajor{Float32}}.
We conclude that we can instantiate and launch customised GEMM kernels easily, without sacrificing flexibility.

\begin{listing}
\begin{minted}{julia}
a = CuArray(rand(Float16, (M, K)))
b = CuArray(rand(Float16, (K, N)))
c = CuArray(rand(Float32, (M, N)))
d = similar(c)

conf = GemmKernels.get_config(
  gemm_shape = (M = M, N = N, K = K),
  operator = Operator.WMMAOp{16, 16, 16},
  global_a_layout = Layout.AlignedColMajor{Float16},
  global_c_layout = Layout.AlignedColMajor{Float32}
                        )

GemmKernels.matmul(
  a, b, c, d, conf;
  transform_regs_to_shared_d = Transform.Elementwise(​x -> max(x, 0))
             )
\end{minted}
    \caption{Matrix product \( D_{ij} = \max(\sum_k A_{ik} \cdot B_{kj} + C_{ij}, 0) \).}
    \label{lst:matmul_relu}
\end{listing}

\subsubsection{BLAS-like interface}%
\label{ssub:blas_like_interface}

\begin{listing}[t]
\begin{minted}{julia}
a = CuArray(rand(Float16, (M, K)))
b = CuArray(rand(Float16, (K, N)))
c = CuArray(rand(Float32, (M, N)))

alpha = rand(Float32)
beta  = rand(Float32)

GemmKernels.BLAS.gemmEx!('N', 'N', alpha, a, b, beta, c)
\end{minted}
    \caption{A matrix product using our BLAS-like interface.}
    \label{lst:matmul_blas}
\end{listing}

\begin{listing}[t]
\begin{minted}{julia}
function library_code(a::CuArray, b::CuArray, c::CuArray)
  # ...
  GemmKernels.BLAS.gemmEx!('N', 'N', alpha, a, b, beta, c)
  # ...
end
\end{minted}
    \caption{Library code making use of our BLAS-like API.}
    \label{lst:library_code}
\end{listing}

\begin{listing}[t]
\begin{minted}{julia}
for f in (:load_a, :load_b, :load_c, :store_d)
  @eval @inline $f(op, ::Type{Layout.Padded{L, P}}, args...) where {L, P} = $f(op, L, args...)
end
\end{minted}
    \caption{Redirecting operator calls for padded layouts to the underlying layout, using Julia's metaprogramming.}
    \label{lst:metaprogramming}
\end{listing}

The interface of the previous section exposes maximal flexibility to the user, but differs from the interface used by \term{cuBLAS}.
We also provide a more familiar BLAS-like interface which can be used if not all flexibility is needed.
This interface supports all operations of \term{cuBLAS}'s \mintinline{text}{gemmEx}, i.e.\ linear scaling and transposition of the input matrices, but, importantly, with support for many more input types.

To ease the transitioning process, this BLAS-like interface has the same signature as \term{CUDA.jl}'s \mintinline{text}{gemmEx} wrapper.
To use our GEMM kernels in existing code, it suffices to replace \mintinline{text}{CUDA.CUBLAS.gemmEx!} by \mintinline{text}{GemmKernels.gemmEx!}.

Using the BLAS interface, there is no need to specify each component manually.
Instead, they are derived from the types of the arguments.
For example, \cref{lst:matmul_blas} calculates the matrix product \( C := \alpha \cdot A \cdot B + \beta \cdot C \).
Based on the combination of the types of the \( A \), \( B \), and \( C \) matrix, our implementation of the BLAS-like interface instantiates a GEMM kernel with \mintinline{text}{operator = Operator.WMMAOp{16, 16, 16}}, \mintinline{text}{global_a_layout = Layout.AlignedColMajor{Float16}}, etc.

Using the BLAS interface in library code allows extending this library with new types, without changing the library code.
For example, the library code in \cref{lst:library_code} is called for GPU arrays, but does not impose any restrictions on the element type.
Using custom element types is as simple as adding support for them in our GEMM framework, and calling the library code with this new type.

\subsection{Discussion}
\label{sub:discussion}

For our GEMM framework design, we have strived for orthogonality of different components.
For example, epilogues and operators only interact via shared memory and can be combined arbitrarily as long as they both support the same shared memory layout.
Transforms use broadcast expressions that work for different data types and array lengths, and can hence be combined with different layouts or params.

Nevertheless, some inevitable coupling between different components remains.
Most prominently, layouts are coupled to epilogues (e.g.\ a bias epilogue can require different logic for row-major and column-major layouts), and to operators (e.g.\ WMMA supporting padded and non-padded layouts).
Luckily, we can reduce the impact on code verbosity and reuse through several features of Julia.
Epilogues can contain layout-agnostic code, and rely on fine-grained method overloading for layout-specific code paths.
Metaprogramming can be used to redirect operator calls for padded layouts to the underlying layout, as illustrated in \Cref{lst:metaprogramming}.

Another point that merits some discussion is the extent to which our APIs and abstractions are Julia-specific, i.e.\ whether or not parts of them can be used for similar APIs in other \glspl{PL}.
While none of our proposed abstractions are Julia-specific, implementations in other \glspl{PL} would suffer from reduced code reuse, or increased overhead and verbosity.
Julia's unique combination of multiple dispatch, type inference, and JIT compilation allows us to compose GEMM operations from different components, without incurring any run time overhead.
Due to the parametric nature of Julia's type system, tile sizes can be moved to the type domain, such that specialised code can be generated per tile size.
Julia's metaprogramming capabilities prove extremely powerful to improve code reuse and reduce code verbosity.

It is also possible to call Julia functions from C code, using Julia's C API.
Since most \glspl{PL} can call C functions, our framework's kernels can be used in C++, Python, C\#, and other high-level \glspl{PL}, and hence also in ML frameworks written in these \glspl{PL}, such as TensorFlow and PyTorch.

Finally, we should discuss the issue of portability. So far, we focused on flexible \glspl{GEMM} for \term{CUDA}-enabled \glspl{GPU}.
Nevertheless, the abstractions in our tiling \gls{API} and flexible \gls{GEMM} \gls{API} are vendor-agnostic, and we expect they can be reused for AMD and Intel GPUs.
More concretely, porting our framework necessitates two changes.
First, our WMMA operator needs to be replaced with an operator using traditional floating point hardware.
Secondly, our template kernel contains \term{CUDA}-specific concepts such as \mintinline{text}{threadIdx}, and hence needs to be ported to \term{OpenCL}.
Code duplication can be avoided using the package \term{KernelAbstractions.jl} that allows writing vendor-agnostic GPU kernels~\cite{churavy2020kernelabstractions}.

\section{Evaluation}%
\label{sec:evaluation}

To evaluate the performance and flexibility of our APIs, we created the necessary components for five GEMM  variants as discussed below.
Run times were measured on an NVIDIA RTX 2080 Ti with NVIDIA Nsight Compute and with \term{BenchmarkTools.jl}, which continues sampling until the standard deviation becomes small enough.
We compare the performance of our kernels to \term{CUTLASS} 2.2, \term{cuTENSOR} 1.3.0 and \term{cuBLAS} 11.2. We set the latter's math mode to \texttt{CUBLAS\_TENSOR\_OP\_MATH} and call \texttt{cublasGemmEx}.
We use \term{CUDA} 11.0, \term{CUDA.jl} 2.0, and Julia 1.5.

\subsection{Mixed-precision GEMM}%
\label{sub:mixed_precision_gemm}

Our first example is a normal mixed-precision GEMM, i.e.\ a computation of the form \( D = A \cdot B + C \).
This operation is directly supported by NVIDIA's \term{cuBLAS} library. To use \glspl{TC} in our GEMM framework, we create an operator that simply calls the correct WMMA functions in our WMMA API for each step in the GEMM's inner loop.

In a GEMM, the \( A \) and \( B \) matrices may be stored in a column-major memory layout (N), or a row-major memory layout (T).
We hence implemented both a \texttt{ColMajor} and \texttt{RowMajor} layout component.
These layouts are suitable for global memory, but lead to inefficient memory accesses in shared memory.
On NVIDIA GPUs, shared memory is split into a set of \emph{memory banks}. Memory accesses to addresses that map to the same bank, so-called bank conflicts, are serialised.

The simplest way to reduce these bank conflicts is to add padding to every column or row, such that the mapping of matrix elements to banks is changed.
To achieve this, we use a \texttt{PaddedLayout} component to store matrices in shared memory. This layout serves as a wrapper for other layouts, e.g.\ \texttt{PaddedLayout\{ColMajor, 8\}} is a column-major layout, where every column is padded by 8 elements.

Out of the three functions associated with layouts, only \mintinline{text}{size} needs to be specialised for each type of padded layout.
This is necessary because padding differs for, e.g.\ row-major layouts vs.\ column-major layouts.
Calls to \mintinline{text}{load} or \mintinline{text}{store} are automatically redirected to the underlying layout.
As a result, supporting padding only required adding 14 lines of source code.
To use padded layouts for a GEMM, it suffices to set, e.g.\ \mintinline{text}{shared_a_layout = Layout.PaddedLayout{Layout.ColMajor, 8}}, similarly to Lines~9--10 in \cref{lst:matmul_relu}.


The epilogue for mixed-precision GEMM simply copies a tile from shared memory to global memory.

\begin{figure}[t]
    \centering
    \includegraphics[width=\linewidth]{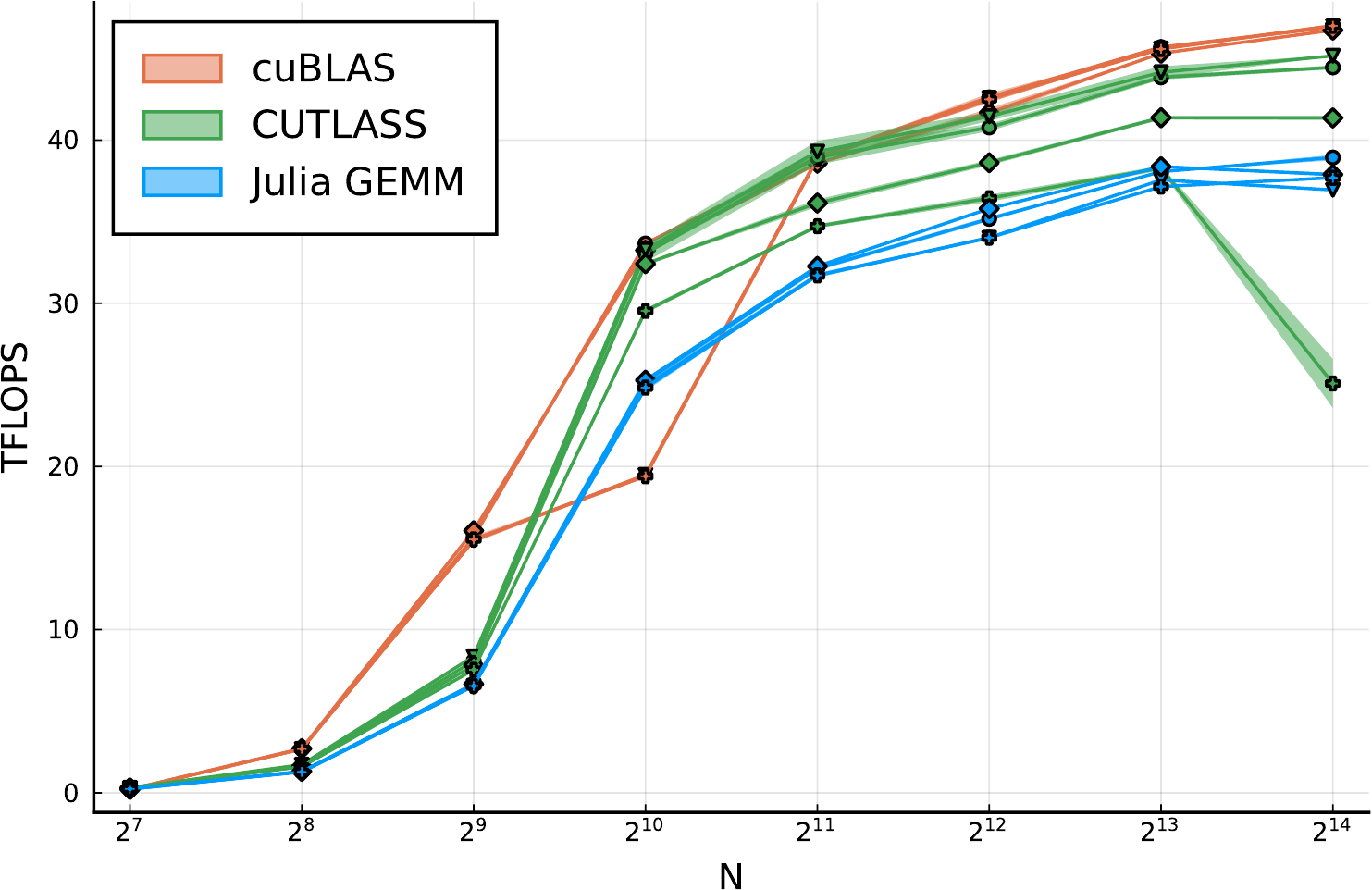}
    \vspace{-0.5cm}
    \caption{Performance of mixed-precision GEMM.}
    \label{fig:performance-normal-gemm}
\end{figure}

\Cref{fig:performance-normal-gemm} compares the performance of our mixed-precision GEMM to \term{CUTLASS} and \term{cuBLAS}.
The different lines and markers represent the four combinations of the data layouts of the \( A \) and \( B \) matrices, the shaded regions represent error margins. We have no explanations for the two anomalous results in the measurements for \term{CUTLASS} and \term{cuBLAS}. They occurred consistently over many experiments.
For the most interesting, larger matrices with N=2048 and more, the performance of our kernels ranges between 82\% and 86\% of \term{cuBLAS}, the best performing library.
We conclude that our kernels achieve reasonably good performance, despite being written completely in Julia. To the best of our knowledge, no existing implementation purely written in a single higher-level \gls{PL} comes close.


The performance difference between our kernel and the state-of-the-art in \term{CUTLASS} and \term{cuBLAS} can be attributed to two factors.
First, our implementation does not yet contain data swizzling to avoid bank conflicts in shared memory. A device-dependent layout is necessary for swizzling optimally for a GPU's shared memory implementation, which we have not yet explored.
Secondly, \term{cuBLAS} does not use WMMA, but accesses \glspl{TC} directly, allowing the use of custom memory layouts in shared memory that perform better than our padded layouts.
Implementing this in our framework necessitates a new layout and a new operator component.
This custom operator would use the \texttt{mma} family of instructions instead of WMMA.
Contrary to WMMA, \texttt{mma} does not have load or store instructions, so distribution of matrix elements to different threads needs to be done explicitly.
Use of the \texttt{mma} operator would hamper portability, though, because some \texttt{mma} instructions are optimised for a particular GPU architecture. Engineering this is future work.

\subsection{Diagonal matrices}%
\label{sub:diagonal_matrices}

Next, we focus on a mixed-precision GEMM where the \( A \) or \( B \) matrix is a diagonal matrix.
In Julia, diagonal matrices are represented using the \texttt{Diagonal} parametric type.
This type acts as a wrapper for a one-dimensional array, which contains the diagonal elements.
References to elements on the diagonal are redirected to loads or stores of this underlying array, whereas references off the diagonal return 0 without performing any actual memory access.

We leveraged Julia's multiple dispatch to provide a GEMM implementation that is specialised for diagonal matrices by means of two optimisations.
First, we replace the layouts from Section~\ref{sub:mixed_precision_gemm} with a \texttt{Diagonal} layout.
Similar to Julia's \texttt{Diagonal} wrapper, this layout simply returns 0 if the accessed element lies off the diagonal, and otherwise accesses an array.
Second, we extended our template GEMM kernel with a customisable predicate that determines for each inner loop iteration if it should be executed or skipped.
By default, this predicate is constant \texttt{true}, but we specialise it for diagonal matrices so that iterations that perform computations on elements off the diagonal are skipped.

\Cref{tab:performance-diagonal} compares the performance of this specialised GEMM kernel with \term{cuBLAS}.
As \term{cuBLAS} does not include GEMM kernels specialised for diagonal matrices, the \term{cuBLAS} version of our code had to materialise the diagonal matrix before calling the standard \term{cuBLAS} GEMM kernel. This is in line with the common practice as discussed in Section~\ref{sub:GEMM_requirements}.
The first optimisation results in a reduction of 89\% in global memory traffic.
The second optimisation reduces the number of \gls{TC} operations by over 95\%.
Together, they lead to a GEMM that is more than 6 times faster than what we can obtain with \term{cuBLAS}'s inflexible kernels.

Adding support for diagonal matrices required adding 23 lines of source code to our existing framework.
We conclude that specialisation of kernels in our framework requires minimal effort, while at least in some cases still obtaining massive performance improvements.

\begin{table}[t]
    \centering
    \caption{Performance of our diagonal matrix GEMM and the equivalent \term{cuBLAS} implementation, for \( N = 4096 \); run times are for 100 iterations.}
    \label{tab:performance-diagonal}
    \begin{tabular}{l|l|l}
        & \textbf{cuBLAS} & \textbf{Ours} \\
        \hline
        \textbf{Run time (ms)} & \( 322.77 \pm 0.08 \)  & \( 50.68 \pm 0.05 \) \\
        \textbf{\#Global mem. accesses} & 3410K per kernel  & 490K per kernel  \\
        \textbf{\#Tensor Core instr.} & 67 109K per kernel  & 3113K per kernel \\
    \end{tabular}
\end{table}

\subsection{Operator fusion}%
\label{sub:operator_fusion}

\term{cuBLAS} fuses linear scaling into its GEMM computation, i.e.\ its GEMM is of the form \( D = \alpha\cdot AB+\beta\cdot C \).
Other computations cannot be fused in \term{cuBLAS}'s kernels and require a separate kernel launch.
We consider two examples of GEMM computations that can exploit operation fusion: custom element-wise operations and adding a bias vector.

In custom element-wise operations, the linear scaling with \( \alpha \) and \( \beta \) is replaced by any arbitrary function.
We implemented this in our GEMM framework using an \texttt{ElementwiseTransform} component.
It has an arbitrary function as a parameter, which is applied to every element.

In bias computations, a one-dimensional bias vector is added to every row of the matrix product.
To add support for bias in our GEMM framework, we created a custom epilogue that loads the bias vector from global memory, and adds it to the matrix product in shared memory, before writing the result back to global memory.

\cref{tab:performance-elop-bias} compares the run times of six GEMMs with and without element-wise operations on input and/or output matrices, and with and without bias vectors.
For the purpose of this experiment, we use additions with a constant and ReLU, a popular function in the domain of ML, as the element-wise operations, but similar results are obtained with other ones that are not supported in \term{CUTLASS}, and for which researchers would have to fall back on \term{cuBLAS} or our solution.
For this reason, we only compare to \term{cuBLAS}.

With \term{cuBLAS}, combining the GEMM with element-wise operations or bias vectors results in additional kernel launches.
While we have fused the operations as much as possible, i.e.\ the application of ReLU on \( D \) and the addition of the bias vector are fused in one kernel, the GEMM and the other operations each still correspond to separate kernels which cannot be fused any further.
By contrast, our framework seamlessly fuses all operations in the GEMM kernel instead.
The effect is clearly visible in the results in the table.
While the standard GEMM in \term{cuBLAS} is faster than with our framework, in line with the results in Section~\ref{sub:mixed_precision_gemm}, our framework catches up as more fusable operations are added.
Whereas the extra operations require almost no additional time in our framework (bias vectors need to be loaded, hence their small cost), each operation requiring a separate kernel costs approximately 10\% in performance with \term{cuBLAS}. Ultimately, our GEMM becomes about 7\% faster.
This clearly illustrates the need for operator fusion, and how effective our approach is.
The fact that our approach fuses the operations seamlessly also implies that any future optimisation of our default implementation of the mixed-precision GEMM, through swizzling or use of \text{mma} instead of WMMA, will automatically benefit GEMMs with fusable operations as well.
Our GEMM will then outperform \term{cuBLAS} even more.

Note that, although our Julia \gls{GEMM} implementation can easily be invoked from within other languages and \gls{ML} frameworks, as discussed in \cref{sub:discussion}, this advantage of fused element-wise operations can only be obtained if they are expressed in Julia, because that fusion depends on the Julia-specific features also discussed in \cref{sub:discussion}.



\begin{table}[t]
    \centering
    \caption{Run times of \term{cuBLAS} that lacks fusion capabilities and our GEMMs that exploit operation fusion, for \( N = 4096 \) and 100 iterations.}
    \label{tab:performance-elop-bias}
    \begin{tabular}{l|l|l}
                                       & \textbf{cuBLAS [ms]}  & \textbf{Ours [ms]}    \\
        \hline
        \textbf{GEMM}                  & \( 260.82 \pm 1.51  \) & \( 312.22 \pm 1.61 \) \\
        \textbf{GEMM + ReLU on D}           & \( 286.20 \pm 1.10 \) & \( 313.00 \pm 1.53 \) \\
        \textbf{GEMM + bias}           & \( 287.72 \pm 1.80 \) & \( 315.56 \pm 1.57 \) \\
        \textbf{GEMM + bias + ReLU on D}    & \( 288.19 \pm 2.15 \) & \(315.54 \pm 1.60 \) \\
        \textbf{GEMM + bias + ReLU on C\,\&\,D}& \( 313.88 \pm 1.22 \) & \( 315.40 \pm 1.95 \) \\
        \textbf{GEMM + bias + ReLU on C\,\&\,D}& \( 340.38 \pm 1.21 \) & \( 316.73 \pm 1.84 \) \\
        \textbf{\hfill + addition operation on A\,\&\,B} & &  \\
    \end{tabular}
\end{table}

\subsection{Complex and dual numbers}%
\label{sub:complex_and_dual_numbers}

Standard mixed-precision GEMMs differ from mixed-precision GEMMs of complex numbers in two ways.
First, the WMMA multiply-accumulate operation in the inner loop is replaced by four WMMA operations: \texttt{A.real * B.real}, \texttt{A.real * B.imag}, \texttt{A.imag * B.real}, and \texttt{A.imag * B.imag}.
Second, complex GEMMs use different memory layouts in global and shared memory.
In global memory, complex matrices are typically stored in an interleaved layout, where the real and imaginary parts are stored contiguously.
This layout is incompatible with WMMA, so in shared memory, we use a split layout instead, where the real and imaginary parts are stored separately.
In our GEMM framework, these two differences correspond to a new operator \texttt{WMMAComplexOp}, and two new layouts \texttt{InterleavedComplex} and \texttt{SplitComplex}, respectively.

Dual numbers differ slightly from complex numbers. The imaginary unit \( i \) is replaced by \( \varepsilon \), and \( \varepsilon^2=0 \) whereas \( i^2=-1 \).
As such, we need an additional \texttt{WMMADualOp} operator component, but we can reuse the split and interleaved layouts we developed for complex matrices.

\begin{figure}[t]
    \centering
    \includegraphics[width=\linewidth]{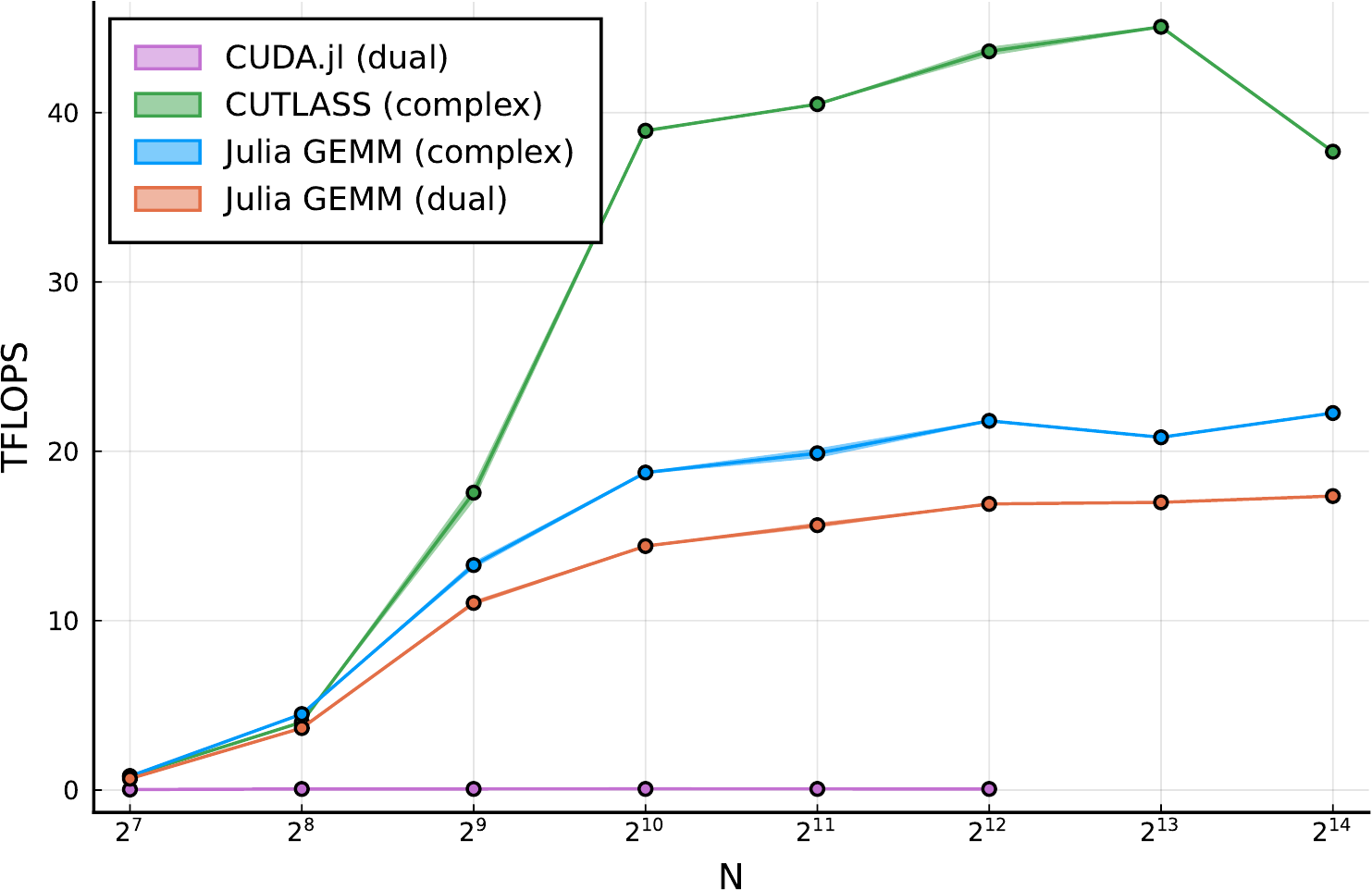}
    \vspace{-0.5cm}
    \caption{Performance of complex and dual GEMMs.}
    \label{fig:performance-complex-dual}
\end{figure}

\cref{fig:performance-complex-dual} shows the performance of four GEMM kernels using complex or dual numbers.
As \term{cuBLAS} supports neither complex numbers using \glspl{TC} nor dual numbers, we did not include it in the comparison. \term{cuBLASLt} does support complex numbers, but uses \term{CUTLASS}'s kernels. \term{CUTLASS} only supports complex numbers. So we compare the performance of our two kernels to \term{CUTLASS} for complex numbers and to the generic \term{CUDA.jl} kernel that is invoked when our API is not used to compute a mixed-precision GEMM on dual numbers. Our complex number implementation achieves a peak performance of 59\% of \term{CUTLASS}'s peak performance.
We conclude that, despite the fact that our kernels are written completely in Julia, and do not contain optimisations specific to complex GEMM, we are still able to reach reasonably good performance.

The difference in performance between our complex GEMM and \term{CUTLASS}'s can partly be attributed to \term{CUTLASS}'s use of a split layout in both global and shared memory, eliminating the overhead of changing layouts in the global-to-shared memory stream.
Our kernels use an interleaved layout in global memory, as this is the format that Julia uses.
This comes with some overhead, but eliminates the need for extra interleaved-to-split and split-to-interleaved kernel launches that would be necessary for \term{CUTLASS}.

While \term{CUDA.jl} contains wrappers for \term{cuBLAS}'s GEMM kernels, it supports only a limited number of data types.
Calling these wrappers with unsupported types, such as dual numbers, falls back to a generic implementation that is many orders of magnitude slower, as is clear from the \term{CUDA.jl} line in \Cref{fig:performance-complex-dual}.

Adding support for complex and dual numbers to our framework required 169 lines of source code, of which 85 are common to both data types.
We conclude that extending our framework with custom data types requires minimal effort, especially compared to the alternative of writing a performant GEMM kernel for these data types from scratch.

\subsection{Tensor Contraction}
\label{sub:tensor_contraction}
Implementing and evaluating fully-optimised general \glspl{TCT} is out-of-scope of this paper. However, we do want to demonstrate that our GEMM functionality does provide the necessary support for common \glspl{TCT}. We hence evaluated the functionality for one \gls{TCT}, namely the first benchmark of the Tensor Contraction Code Generator benchmarks~\cite{springer2018design}. In Einstein notation, this \gls{TCT} is written as $D_{abc} = A_{bda} \cdot B_{dc}$.

Using our GEMM APIs, we created custom global memory layouts for matrices $A$ to $D$. The layout for $C$ always returns zero without performing loads to global memory. The layouts for $A$, $B$, and $D$ provide the fused transpositions of GETT~\cite{springer2016design}. As the contraction index $d$'s position is almost perfect for an NN-GEMM, the $A$, $B$, and $D$ layouts started from the  \mintinline{text}{AlignedColMajor{T}} layouts. Custom mappings from elements in global memory to shared memory were implemented such that GEMM in shared memory is equivalent with \gls{TCT} in global memory. We did not extensively search for optimal memory layouts, but simply tried to use as many stride-1 accesses in global memory as possible, to facilitate vectorisation and global memory coalescing. As an example, the $A$ layout code is shown in \cref{lst:tc_layout}. With this layout, $A_{bda}$ in global memory becomes $A_{bad}$ in shared memory, which is then interpreted as the 2D matrix $A_{mk}$, by mapping of dimensions $(ba) \rightarrow m$ and $d \rightarrow k$. $B$ is $B_{dc}$ in global memory and stays the same in shared memory, with mapping $d \rightarrow k$ and $c \rightarrow n$. The GEMM in shared memory then computes $A_{mk} B_{kn} = A_{bad} B_{dc} = D_{bac} $, and the custom $D$ layout ensures shuffling is performed to $D_{abc}$ while copying data from shared to global memory.

\begin{listing}
\begin{minted}{julia}
abstract type LayoutA{T} <: Layout.AlignedColMajor{T} end

@inline function Layout.load(::Type{LayoutA{T}},​ workspace, tile::Tile{size}) where {T, size}
  NUMEL = 16 ÷ sizeof(T)

  M = tile.base.M + tile.offset.M
  K = tile.base.K + tile.offset.K
  d = K
  a = M ÷ Base.size(workspace, 1)
  b = M % Base.size(workspace, 1)

  offset = 1 + b + d * Base.size(workspace, 1) +​ a * Base.size(workspace, 1) * Base.size(workspace, 2)
  Layout.vloada(Layout.Vec{NUMEL, T}, pointer(workspace),​ offset)
end
\end{minted}
    \caption{Layout for tensor \( A \).}
    \label{lst:tc_layout}
\end{listing}

We compare the performance of our implementation to that in \term{cuTENSOR}, a hand-optimised tensor library from NVIDIA. We observed that \term{cuTENSOR} by default does not use the GETT approach for the evaluated \gls{TCT}, as it executes two kernels, similar to the TTGT approach~\cite{solomonik2013cyclops,bader2006algorithm}. We can however force it to use the GETT approach as well.

\begin{table}[t]
    \centering
    \caption{Performance of \gls{TCT} and the equivalent \term{cuTENSOR} implementations, for \( N_a = 64, N_b = 32, N_c = 2048, N_d = 2048 \); run times are in \si{\micro\second}.}
    \label{tab:performance_ct}
    \begin{tabular}{c|c|c}
        \textbf{cuTENSOR TTGT} & \textbf{cuTENSOR GETT} & \textbf{Ours} \\
        \hline
        \( 351.14 \pm 2.46 \)  & \( 1238.81 \pm 9.39 \) & \( 433.82 \pm 5.26 \) \\
    \end{tabular}
\end{table}

\cref{tab:performance_ct} lists the performance results.
Our kernels achieve 81\% of the performance of \term{cuTENSOR} TTGT, the best performing configuration.
Interestingly, \term{cuTENSOR}'s GETT kernel performs \( 3.5\times \) worse than TTGT, and we outperform it by a factor of \( 2.9 \).

The \( B \) and \( D \) layouts look similar to \cref{lst:tc_layout}; the \( C \) layout always returns \( 0 \) and elides the indexing logic and memory operations, but otherwise has the same structure.
In total, these four layouts add up to 58 lines of code.

We conclude that our framework can support (at least some variants of) \glspl{TCT}, and achieves performance in the same ballpark as that of state-of-the-art kernels in \term{cuTENSOR}, while requiring minimal programming effort.



\section{Related work}%
\label{sec:related}
In Section~\ref{sub:GEMM_requirements}, we already discussed the need for flexible and performant GEMMs on GPUs in various domains. As discussed in Section~\ref{sub:tensor_cores}, several libraries exist to deliver the performance. We compared the performance of our GEMM API with some of them in Section~\ref{sec:evaluation}.

NVIDIA's \term{CUTLASS} template library contains components to instantiate performant GEMMs on GPUs~\cite{nvidia2020cutlass}.
As it is written in C++, it does not solve the two-language problem and its impact on programmer productivity.
It did serve, however, as an inspiration for the abstractions in our GEMM API.
For example, \term{CUTLASS} also has the notion of layouts that map logical indices to physical memory offsets, and epilogues for custom post-processing.
Some components have a slightly different purpose, however.
In \term{CUTLASS}, transforms only apply to the global-to-shared memory stream of the \( A \) and \( B \) matrices.
Element-wise transforms on the resultant matrix are handled in a custom epilogue.
Adding support for custom transformations requires significant effort, as \term{CUTLASS} epilogues are typically 150--200 lines long.
In our GEMM API, transforms are applied after every load and before every store, ensuring that element-wise operations to the input and resultant matrices can be applied more easily and consistently.

The \term{CUTLASS} codebase is extensive and contains many components, making it more difficult for end users to get started extending it.
Each GEMM typically involves quite a few layered template instantiations, impacting code comprehension.
Most templates are heavily specialised for different memory layouts, computations, etc., reducing orthogonality and code reuse.
For example, \term{CUTLASS} contains different epilogues for GEMMs exploiting \glspl{TC} and GEMMs using FPUs. Our GEMM API abstractions offer better separation of concerns and hence more reusability.

Like our GEMM API, \term{CUTLASS} contains both a BLAS-like interface, and an interface exposing all its flexibility.
Launching a \term{CUTLASS} kernel using the latter requires more boilerplate compared to our approach in \cref{lst:matmul_relu}, e.g.\ because \term{CUTLASS} users need to explicitly allocate a workspace that is used internally in \term{CUTLASS}.

NVIDIA's \term{cuBLASLt} is a lightweight BLAS library dedicated to GEMM, with a more flexible API than \term{cuBLAS}.
This flexibility comes in the form of support for more matrix layouts, input and compute data types, and algorithmic implementations.
It is available on \term{CUDA} 10.1 or later.

Like our GEMM API, launching a kernel in \term{cuBLASLt} is also a two-step process.
First, a ``plan'' must be created that determines the options for the GEMM computation.
Second, this plan is used to launch one or more GEMM kernels.

\term{cuBLASLt} features concepts similar to \term{CUTLASS}, such as epilogues that post-process the resultant matrix and layouts describing how matrices are stored in memory.
Each of these corresponds to an enumeration that lists the legal values, hence limiting flexibility.
For example, it only includes bias and ReLU as possible epilogues, and offers no support for custom layouts such as diagonal matrices, or custom data types such as dual numbers.
While \term{cuBLASLt}'s interface is an improvement over that of \term{cuBLAS}, its closed-source nature still results in limited extensibility.

\term{BLIS} is a framework that facilitates the instantiation of an entire \term{BLAS} library for new architectures, and hence has a larger scope than just \gls{GEMM}~\cite{vanzee2015blis}.
It achieves this by rephrasing all \term{BLAS} operations in terms of a limited set of kernels.
Their focus is on CPUs rather than GPUs, however.
Similar to our \gls{GEMM} kernel, \term{BLIS} contains a set components that can be reused for new \term{BLAS}-like operations.
\term{BLIS}'s \gls{GEMM} kernels offer support for more memory layouts and data types than traditional \term{BLAS} libraries.
Nevertheless, \term{BLIS}'s flexibility is mainly aimed at developers of the \term{BLIS} library, instead of its users.
For example, extending \term{BLIS} with support for complex numbers required significant effort, that warranted a separate paper describing its implementation~\cite{vanzee2020implementing}.

Section~\ref{sec:tiling} presented a tiling API that allows programmers to coordinate memory transfers to improve data locality.
Automated tools based on polyhedral optimisation exist that can automatically generate tiled code from nested loops~\cite{bondhugula2008practical,verdoolaege2013polyhedral,di2012automatic}.
Basic approaches only reorder memory accesses, but more advanced ones can also exploit parallelism.
For example, \term{Polly} can exploit inter-tile parallelism using the \term{OpenMP} interface~\cite{grosser2012polly}.
The framework by Baskaran et al.\ is even capable of automatically adding padding for shared memory accesses to reduce bank conflicts~\cite{baskaran2008compiler}.

Recent work by Bondhugula uses the polyhedral utilities in \term{MLIR} to generate performant GEMM kernels~\cite{bondhugula2020high}.
His approach achieves a performance that is within 9\% of state-of-the-art CPU \glspl{GEMM} in \term{BLIS} and \term{MKL}.
It still offers limited flexibility, however.
Incorporating domain-specific optimisations, such as diagonal matrices, is significantly harder than in our approach, as it requires adaptations to the \term{MLIR} code base and/or \term{TableGen} rules.

The \term{Diesel} \gls{DSL} uses polyhedral techniques to compile high-level expressions to performant \gls{GPU} kernels~\cite{elango2018diesel}.
Bhaskaracharya et al.\ extend \term{Diesel} with support for \glspl{TC} and fused kernels that combine matrix multiplications with ReLU and bias~\cite{bhaskaracharya2020automatic}.
Their work focuses only on Volta \glspl{TC}, however, and does not address other forms of GEMM flexibility such as support for more complex data types.
Additionally, only a limited number of element-wise operations are supported.

\term{Halide}, a \gls{DSL} for image processing, was also extended with \gls{TC} support by Sioutas et al.~\cite{sioutas2020programming}.
Their approach features a fixed kernel skeleton, similar to our template GEMM kernel.
Their kernel also makes use of WMMA, and achieves performance results similar to ours.
It still has limited flexibility, however. For example, it offers no support for complex data types, and only handles one combination of memory layouts for the \( A \) and \( B \) matrices, which necessitates explicit transposition kernels.

Note that \term{BLIS}, the polyhedral techniques (in so far as they support GPUs), \term{Diesel} and \term{Halide} all involve statically compiled, statically typed \glspl{PL}. For the general-purpose \glspl{PL}, this by itself foregoes the productivity advantages of rapid-prototyping \glspl{PL} such as Julia or Python. For the \glspl{DSL}, it implies that code reuse across domains is limited. Those solutions also by construction do not consider flexibility beyond the data types, layouts, and operations typical for their domains.  Our solution does not suffer from these drawbacks, yet obtains performance in the same ballpark.

\section{Availability}
\label{sec:availability}

Our contributions are open source and available in the relevant GitHub repositories.
Support for Tensor Cores using WMMA was merged into \term{CUDA.jl}, and is available in the latest stable version.
The required adaptations to the Julia compiler were sent to the developers, and have been merged upstream.
Our tiling and flexible GEMM APIs are bundled in one Julia package \term{GemmKernels.jl}.
This package is available at \url{https://github.com/thomasfaingnaert/GemmKernels.jl}, and can easily be installed using Julia's built-in package manager.
It contains all instantiations of our API abstractions of all experiments and listings in this paper, ready for out-of-the-box re-use.

\section{Conclusions and future work}%
\label{sec:conclusions}

In this paper, we first presented tiling abstractions with which programmers can use tiling techniques, which are necessary to achieve high performance for many computations, at a high level of abstraction.

We then discussed a flexible GEMM API where the kernel consists of a set of orthogonal components.
Each of these components corresponds to a set of Julia functions that can be specialised for different GEMM variants.
We demonstrated the flexibility of this approach by instantiating the necessary components for 5 variants of GEMM computations: a normal mixed-precision GEMM, computations using diagonal matrices, computations exploiting operator fusion, GEMMs on complex and dual numbers, and a tensor contraction.
We argued how specific features of the Julia compiler, such as multiple dispatch, type inference, and just-ahead-of-time compilation, allow for this flexibility without run-time overhead.

An experimental evaluation showed that the performance of our GEMM kernels written entirely in Julia is in the same ballpark as, and in some cases even exceeds, the state of the art in the manually tuned \term{cuBLAS}, \term{CUTLASS}, and \term{cuTENSOR}.

We presented two interfaces to use our flexible GEMM API: a fully-featured interface and a \term{BLAS}-like interface.
The former exposes the full flexibility of our framework, the latter extends \term{BLAS}'s GEMM with support for more data types such as dual numbers. We demonstrated our APIs for \term{CUDA}-enabled \glspl{GPU}, but our abstractions are vendor-agnostic and can be ported to other GPU architectures.

In the future, we plan to port our framework to other GPUs such as those of AMD and Intel, and to add support for the \mintinline{text}{mma} family of instructions as well as data swizzling, as used in \term{CUTLASS} and \term{cuBLAS}, to improve performance.
At the moment, the matrix inputs to our kernels must be zero-padded such that their size is a multiple of the GEMM tile sizes.
The engineering required to support arbitrary matrix dimensions, e.g.\ through the use of predicated memory accesses as done by \term{CUTLASS}, is also future work.


\section*{Acknowledgements}
This work was funded by the Research
Foundation Flanders (Fonds voor Wetenschappelijk Onderzoek), grant number 3G051318.

\bibliographystyle{IEEEtran}
\bibliography{main}

\begin{thebibliography}{10}
\providecommand{\url}[1]{#1}
\csname url@samestyle\endcsname
\providecommand{\newblock}{\relax}
\providecommand{\bibinfo}[2]{#2}
\providecommand{\BIBentrySTDinterwordspacing}{\spaceskip=0pt\relax}
\providecommand{\BIBentryALTinterwordstretchfactor}{4}
\providecommand{\BIBentryALTinterwordspacing}{\spaceskip=\fontdimen2\font plus
\BIBentryALTinterwordstretchfactor\fontdimen3\font minus
  \fontdimen4\font\relax}
\providecommand{\BIBforeignlanguage}[2]{{%
\expandafter\ifx\csname l@#1\endcsname\relax
\typeout{** WARNING: IEEEtran.bst: No hyphenation pattern has been}%
\typeout{** loaded for the language `#1'. Using the pattern for}%
\typeout{** the default language instead.}%
\else
\language=\csname l@#1\endcsname
\fi
#2}}
\providecommand{\BIBdecl}{\relax}
\BIBdecl

\bibitem{blas2017home}
{BLAS contributors}. (2017) {BLAS (Basic Linear Algebra Subprograms)}.

\bibitem{abdelfattah2019towards}
A.~{Abdelfattah}, S.~{Tomov}, and J.~{Dongarra}, ``Towards half-precision
  computation for complex matrices: A case study for mixed precision solvers on
  {GPUs},'' in \emph{IEEE/ACM 10th Workshop on Latest Advances in Scalable
  Algorithms for Large-Scale Systems}, 2019, pp. 17--24.

\bibitem{haidar2018harnessing-iterative-refinement}
A.~Haidar, S.~Tomov, J.~Dongarra, and N.~J. Higham, ``Harnessing {GPU} {Tensor
  Cores} for fast {FP16} arithmetic to speed up mixed-precision iterative
  refinement solvers,'' in \emph{Proceedings of the International Conference
  for High Performance Computing, Networking, Storage, and Analysis}, ser. SC
  ’18.\hskip 1em plus 0.5em minus 0.4em\relax IEEE Press, 2018.

\bibitem{ichimura2018fast}
T.~Ichimura, K.~Fujita, T.~Yamaguchi, A.~Naruse, J.~C. Wells, T.~C. Schulthess,
  T.~P. Straatsma, C.~J. Zimmer, M.~Martinasso, K.~Nakajima, M.~Hori, and
  L.~Maddegedara, ``A fast scalable implicit solver for nonlinear
  time-evolution earthquake city problem on low-ordered unstructured finite
  elements with artificial intelligence and transprecision computing,'' in
  \emph{Proc.\ Int'l Conference for High Performance Computing, Networking,
  Storage, and Analysis}, 2018.

\bibitem{haidar2018harnessing-hpc-scientific-applications}
A.~Haidar, H.~Bayraktar, S.~Tomov, and J.~Dongarra, ``Harnessing {Tensor Cores}
  {FP16} arithmetic to accelerate linear solvers and {HPC} scientific
  applications,'' 2018, nVIDIA GPU Technology Conference.

\bibitem{mehta2019getting}
V.~Mehta, ``Getting started with {Tensor Cores} in {HPC},'' 2019, nVIDIA GPU
  Technology Conference.

\bibitem{yan2020demystifying}
D.~Yan, W.~Wang, and X.~Chu, ``Demystifying {Tensor Cores} to optimize
  half-precision matrix multiply,'' in \emph{Proc.\ 34th IEEE International
  Parallel and Distributed Processing Symposium}, 2020.

\bibitem{nvidia2020deep}
{NVIDIA}. (2019, 6) Deep learning performance guide.

\bibitem{nvidia2020v100}
------. (2020) {NVIDIA V100}.

\bibitem{barham2019machine}
P.~Barham and M.~Isard, ``Machine learning systems are stuck in a rut,'' in
  \emph{Proc.\ Workshop on Hot Topics in Operating Systems}, 2019, pp.
  177–--183.

\bibitem{rink2018cfdlang}
N.~Rink, A.~Susungi, J.~Castrillón, J.~Stiller, and C.~Tadonki, ``{CFDlang}:
  High-level code generation for high-order methods in fluid dynamics,'' in
  \emph{Real World Domain Specific Languages Workshop 2018}, 02 2018, pp.
  1--10.

\bibitem{poya2017ahighperformance}
R.~Poya, A.~J. Gil, and R.~Ortigosa, ``{A high performance data parallel tensor
  contraction framework: Application to coupled electro-mechanics},''
  \emph{Computer Physics Communications}, vol. 216, pp. 35--52, 2017.

\bibitem{auer2006automatic}
A.~Auer, G.~Baumgartner, D.~Bernholdt, A.~Bibireata, V.~Choppella, D.~Cociorva,
  G.~Xiaoyang, R.~Harrison, S.~Krishnamoorthy, S.~Krishnan, C.-C. Lam, Q.~Lu,
  M.~Nooijen, R.~Pitzer, J.~Ramanujam, P.~Sadayappan, and A.~Sibiryakov,
  ``Automatic code generation for many-body electronic structure methods: The
  tensor contraction engine,'' \emph{Molecular Physics}, vol. 104, 01 2006.

\bibitem{springer2017landscape}
P.~Springer and P.~Bientinesi, ``The landscape of high-performance tensor
  contractions,'' in \emph{Workshop on Batched, Reproducible, and Reduced
  Precision BLAS}, 2017.

\bibitem{nelson2015generating}
T.~{Nelson}, A.~{Rivera}, P.~{Balaprakash}, M.~{Hall}, P.~D. {Hovland},
  E.~{Jessup}, and B.~{Norris}, ``Generating efficient tensor contractions for
  {GPUs},'' in \emph{2015 44th International Conference on Parallel
  Processing}, 2015, pp. 969--978.

\bibitem{springer2018design}
P.~Springer and P.~Bientinesi, ``Design of a high-performance {GEMM}-like
  tensor--tensor multiplication,'' \emph{ACM Transactions on Mathematical
  Software (TOMS)}, vol.~44, no.~3, pp. 1--29, 2018.

\bibitem{napoli2014towards}
E.~D. Napoli, D.~Fabregat-Traver, G.~Quintana-Ort{\'{i}}, and P.~Bientinesi,
  ``{Towards an efficient use of the {BLAS} library for multilinear tensor
  contractions},'' \emph{Applied Mathematics and Computation}, vol. 235, pp.
  454--468, 2014.

\bibitem{li2015aninput}
J.~{Li}, C.~{Battaglino}, I.~{Perros}, J.~{Sun}, and R.~{Vuduc}, ``An
  input-adaptive and in-place approach to dense tensor-times-matrix multiply,''
  in \emph{Proc.\ Int'l Conference for High Performance Computing, Networking,
  Storage and Analysis}, 2015, pp. 1--12.

\bibitem{solomonik2013cyclops}
E.~{Solomonik}, D.~{Matthews}, J.~{Hammond}, and J.~{Demmel}, ``Cyclops tensor
  framework: Reducing communication and eliminating load imbalance in massively
  parallel contractions,'' in \emph{27th Int'l Symposium on Parallel and
  Distributed Processing}, 2013, pp. 813--824.

\bibitem{bader2006algorithm}
B.~W. Bader and T.~G. Kolda, ``Algorithm 862: {MATLAB} tensor classes for fast
  algorithm prototyping,'' \emph{ACM Transactions on Mathematical Software},
  vol.~32, no.~4, pp. 635--653, December 2006.

\bibitem{kim2019acode}
J.~Kim, A.~Sukumaran-Rajam, V.~Thumma, S.~Krishnamoorthy, A.~Panyala, L.-N.
  Pouchet, A.~Rountev, and P.~Sadayappan, ``A code generator for
  high-performance tensor contractions on {GPUs},'' in \emph{Proc.\ IEEE/ACM
  Int'l Symposium on Code Generation and Optimization}, 2019, p. 85–95.

\bibitem{matthews2018high}
D.~A. Matthews, ``High-performance tensor contraction without transposition,''
  \emph{SIAM Journal on Scientific Computing}, vol.~40, no.~1, pp. C1--C24,
  2018.

\bibitem{psarras2021landscape}
C.~Psarras, L.~Karlsson, J.~Li, and P.~Bientinesi, ``The landscape of software
  for tensor computations,'' 2021.

\bibitem{nvidia2020cuda}
{NVIDIA}. (2020) {CUDA} {C++} programming guide.

\bibitem{khronos2020opencl}
{Khronos Group}. (2020) {OpenCL}: An open standard for parallel programming of
  heterogeneous systems.

\bibitem{julia2020home}
{JuliaLang.org}. (2020) The {Julia} language.

\bibitem{julia2020benchmarks}
------. (2020) {Julia} micro-benchmarks.

\bibitem{besard2019effective}
T.~{Besard}, C.~{Foket}, and B.~{De Sutter}, ``Effective extensible
  programming: Unleashing {Julia} on {GPUs},'' \emph{IEEE Transactions on
  Parallel and Distributed Systems}, vol.~30, no.~4, pp. 827--841, 2019.

\bibitem{besard2019rapid}
T.~Besard, V.~Churavy, A.~Edelman, and B.~{De Sutter}, ``Rapid software
  prototyping for heterogeneous and distributed platforms,'' \emph{Advances in
  Engineering Software}, vol. 132, pp. 29 -- 46, 2019.

\bibitem{julia2020docs}
{JuliaLang.org}. (2020) The {Julia} language official documentation.

\bibitem{llvm2020home}
{LLVM contributors}. (2020) The {LLVM} compiler infrastructure project.

\bibitem{churavy2020gpuifyloops}
V.~Churavy. (2020) {GPUifyLoops.jl}: Support for writing loop-based code that
  executes both on {CPU} and {GPU}.

\bibitem{hinston2018matrix}
G.~Hinton, S.~Sabour, and N.~Frosst, ``Matrix capsules with {EM} routing,'' in
  \emph{International Conference on Learning Representations}, 2018.

\bibitem{abadi2016tensorflow}
M.~Abadi, P.~Barham, J.~Chen, Z.~Chen, A.~Davis, J.~Dean, M.~Devin,
  S.~Ghemawat, G.~Irving, M.~Isard, M.~Kudlur, J.~Levenberg, R.~Monga,
  S.~Moore, D.~G. Murray, B.~Steiner, P.~Tucker, V.~Vasudevan, P.~Warden,
  M.~Wicke, Y.~Yu, and X.~Zheng, ``Tensorflow: A system for large-scale machine
  learning,'' in \emph{Proc.\ 12th USENIX Symposium on Operating Systems Design
  and Implementation (OSDI)}, 2016, pp. 265--283.

\bibitem{paszke2019pytorch}
A.~Paszke, S.~Gross, F.~Massa, A.~Lerer, J.~Bradbury, G.~Chanan, T.~Killeen,
  Z.~Lin, N.~Gimelshein, L.~Antiga, A.~Desmaison, A.~Kopf, E.~Yang, Z.~DeVito,
  M.~Raison, A.~Tejani, S.~Chilamkurthy, B.~Steiner, L.~Fang, J.~Bai, and
  S.~Chintala, ``{PyTorch}: An imperative style, high-performance deep learning
  library,'' in \emph{Advances in Neural Information Processing Systems 32},
  2019, pp. 8024--8035.

\bibitem{apra2014efficient}
E.~{Aprà}, M.~{Klemm}, and K.~{Kowalski}, ``Efficient implementation of
  many-body quantum chemical methods on the {Intel® Xeon Phi} coprocessor,''
  in \emph{Proc.\ Int'l Conference for High Performance Computing, Networking,
  Storage and Analysis}, 2014, pp. 674--684.

\bibitem{ma2011gpu}
W.~Ma, S.~Krishnamoorthy, O.~Villa, and K.~Kowalski, ``{GPU-Based
  Implementations of the Noniterative Regularized-CCSD(T) Corrections:
  Applications to Strongly Correlated Systems},'' \emph{Journal of Chemical
  Theory and Computation}, vol.~7, no.~5, pp. 1316--1327, 2011.

\bibitem{springer2016design}
P.~Springer and P.~Bientinesi, ``Design of a high-performance {GEMM}-like
  tensor-tensor multiplication,'' 2016.

\bibitem{revels2016autodiff}
J.~{Revels}, M.~{Lubin}, and T.~{Papamarkou}, ``Forward-mode automatic
  differentiation in {J}ulia,'' \emph{arXiv:1607.07892 [cs.MS]}, 2016.

\bibitem{churavy2020kernelabstractions}
\BIBentryALTinterwordspacing
V.~Churavy. (2020) {KernelAbstractions.jl}: Heterogeneous programming in
  {Julia}. [Online]. Available:
  \url{https://github.com/JuliaGPU/KernelAbstractions.jl}
\BIBentrySTDinterwordspacing

\bibitem{nvidia2020cutlass}
{NVIDIA}. (2020) {CUTLASS}: {CUDA} templates for linear algebra subroutines.

\bibitem{vanzee2015blis}
F.~G. Van~Zee and R.~A. van~de Geijn, ``{BLIS}: A framework for rapidly
  instantiating {BLAS} functionality,'' \emph{ACM Trans. Math. Softw.},
  vol.~41, no.~3, Jun. 2015.

\bibitem{vanzee2020implementing}
F.~G. Van~Zee, ``Implementing high-performance complex matrix multiplication
  via the {1M} method,'' \emph{SIAM Journal on Scientific Computing}, vol.~42,
  no.~5, pp. C221--C244, 2020.

\bibitem{bondhugula2008practical}
U.~Bondhugula, A.~Hartono, J.~Ramanujam, and P.~Sadayappan, ``A practical
  automatic polyhedral parallelizer and locality optimizer,'' in \emph{Proc.\
  29th ACM SIGPLAN Conference on Programming Language Design and
  Implementation}, 2008, pp. 101--113.

\bibitem{verdoolaege2013polyhedral}
S.~Verdoolaege, J.~Carlos~Juega, A.~Cohen, J.~Ignacio~Gomez, C.~Tenllado, and
  F.~Catthoor, ``Polyhedral parallel code generation for {CUDA},'' \emph{ACM
  Transactions on Architecture and Code Optimization (TACO)}, vol.~9, no.~4,
  pp. 1--23, 2013.

\bibitem{di2012automatic}
P.~Di, D.~Ye, Y.~Su, Y.~Sui, and J.~Xue, ``Automatic parallelization of tiled
  loop nests with enhanced fine-grained parallelism on {GPUs},'' in
  \emph{Proc.\ 41st Int'l Conference on Parallel Processing}, 2012, pp.
  350--359.

\bibitem{grosser2012polly}
T.~Grosser, A.~Groesslinger, and C.~Lengauer, ``Polly—performing polyhedral
  optimizations on a low-level intermediate representation,'' \emph{Parallel
  Processing Letters}, vol.~22, no.~04, p. 1250010, 2012.

\bibitem{baskaran2008compiler}
M.~M. Baskaran, U.~Bondhugula, S.~Krishnamoorthy, J.~Ramanujam, A.~Rountev, and
  P.~Sadayappan, ``A compiler framework for optimization of affine loop nests
  for {GPGPUs},'' in \emph{Proc.\ 22nd Annual Int'l Conference on
  Supercomputing}, 2008, pp. 225--234.

\bibitem{bondhugula2020high}
U.~Bondhugula, ``High performance code generation in {MLIR}: An early case
  study with {GEMM},'' \emph{preprint arXiv:2003.00532}, 2020.

\bibitem{elango2018diesel}
V.~Elango, N.~Rubin, M.~Ravishankar, H.~Sandanagobalane, and V.~Grover,
  ``Diesel: {DSL} for linear algebra and neural net computations on {GPUs},''
  in \emph{Proc.\ 2nd ACM SIGPLAN Int'l Workshop on Machine Learning and
  Programming Languages}, 2018, pp. 42--51.

\bibitem{bhaskaracharya2020automatic}
S.~G. Bhaskaracharya, J.~Demouth, and V.~Grover, ``Automatic kernel generation
  for {Volta} {Tensor Cores},'' \emph{arXiv preprint arXiv:2006.12645}, 2020.

\bibitem{sioutas2020programming}
S.~Sioutas, S.~Stuijk, T.~Basten, L.~Somers, and H.~Corporaal, ``Programming
  tensor cores from an image processing {DSL},'' in \emph{Proc.\ 23th Int'l
  Workshop on Software and Compilers for Embedded Systems}, 2020, pp. 36--41.

\end{thebibliography}

\begin{IEEEbiography}[{\includegraphics[height=1.25in]{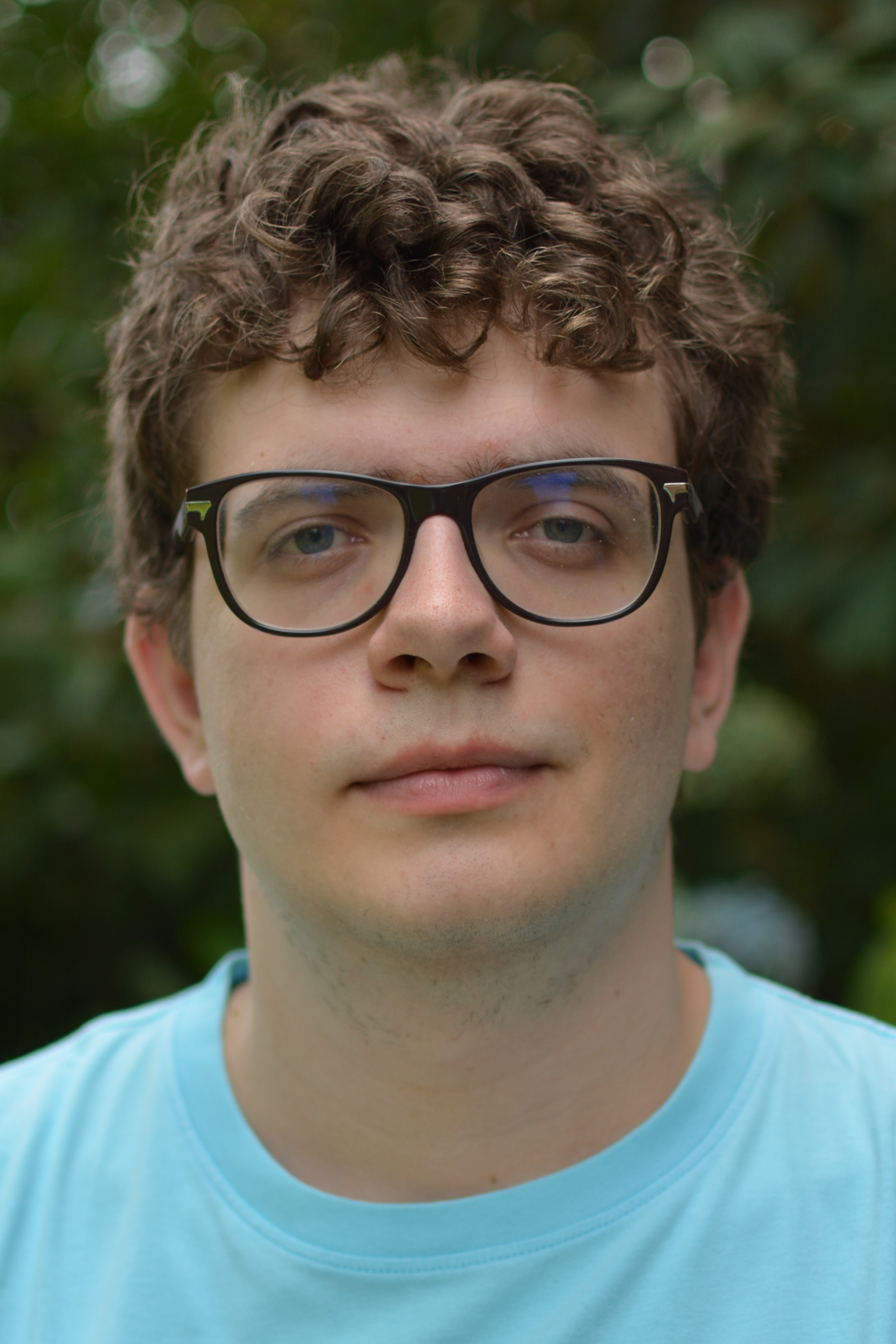}}]
{Thomas Faingnaert} is a PhD student at Ghent University in the Computer Systems Lab.
He obtained his MSc degree in Computer Science Engineering from Ghent University's Faculty of Engineering and Architecture in 2020.
His research focuses on software protection, and high-level abstractions for GPU programming in Julia.
\end{IEEEbiography}
\vspace*{-2\baselineskip}
\begin{IEEEbiography}[{\includegraphics[height=1.25in]{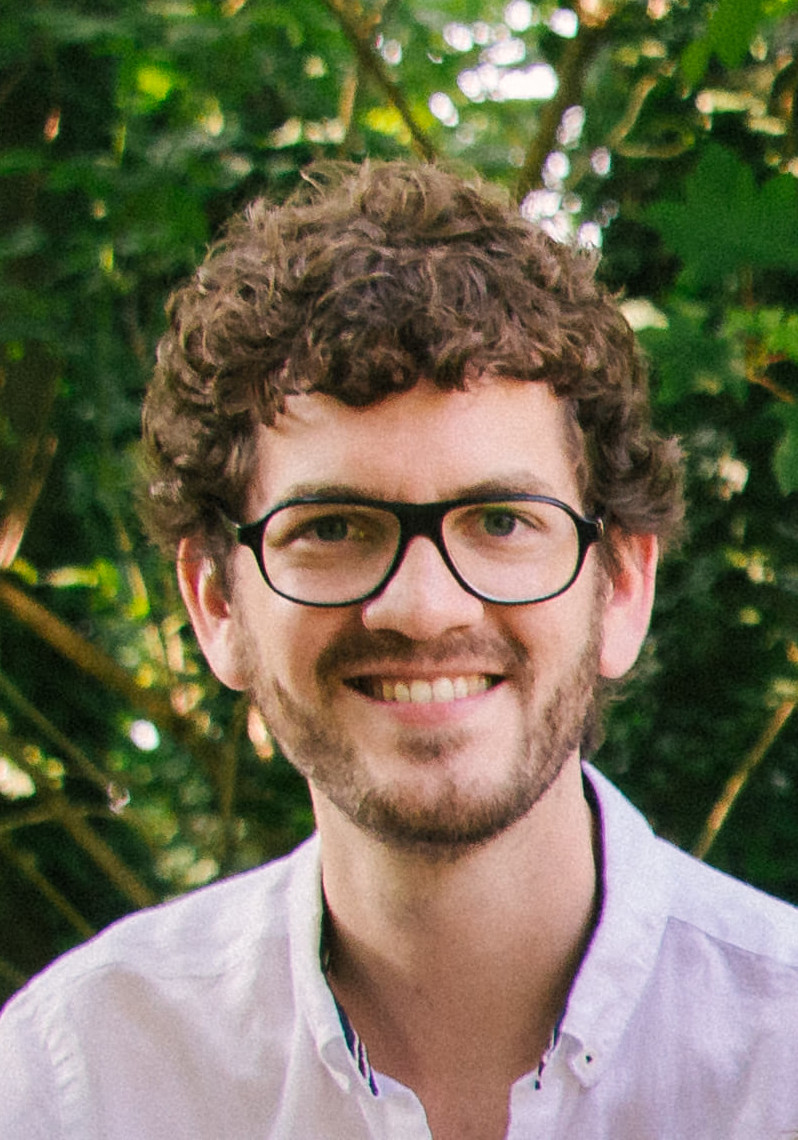}}]
{Tim Besard} is a software engineer at Julia Computing. He obtained his MSc in Computer
Engineering from University College Ghent in 2011, and his PhD in Computer Science
Engineering from Ghent University in 2019. He is currently the lead maintainer of several
GPU back-ends for the Julia programming language.
\end{IEEEbiography}
\vspace*{-2\baselineskip}
\begin{IEEEbiography}[{\includegraphics[height=1.25in]{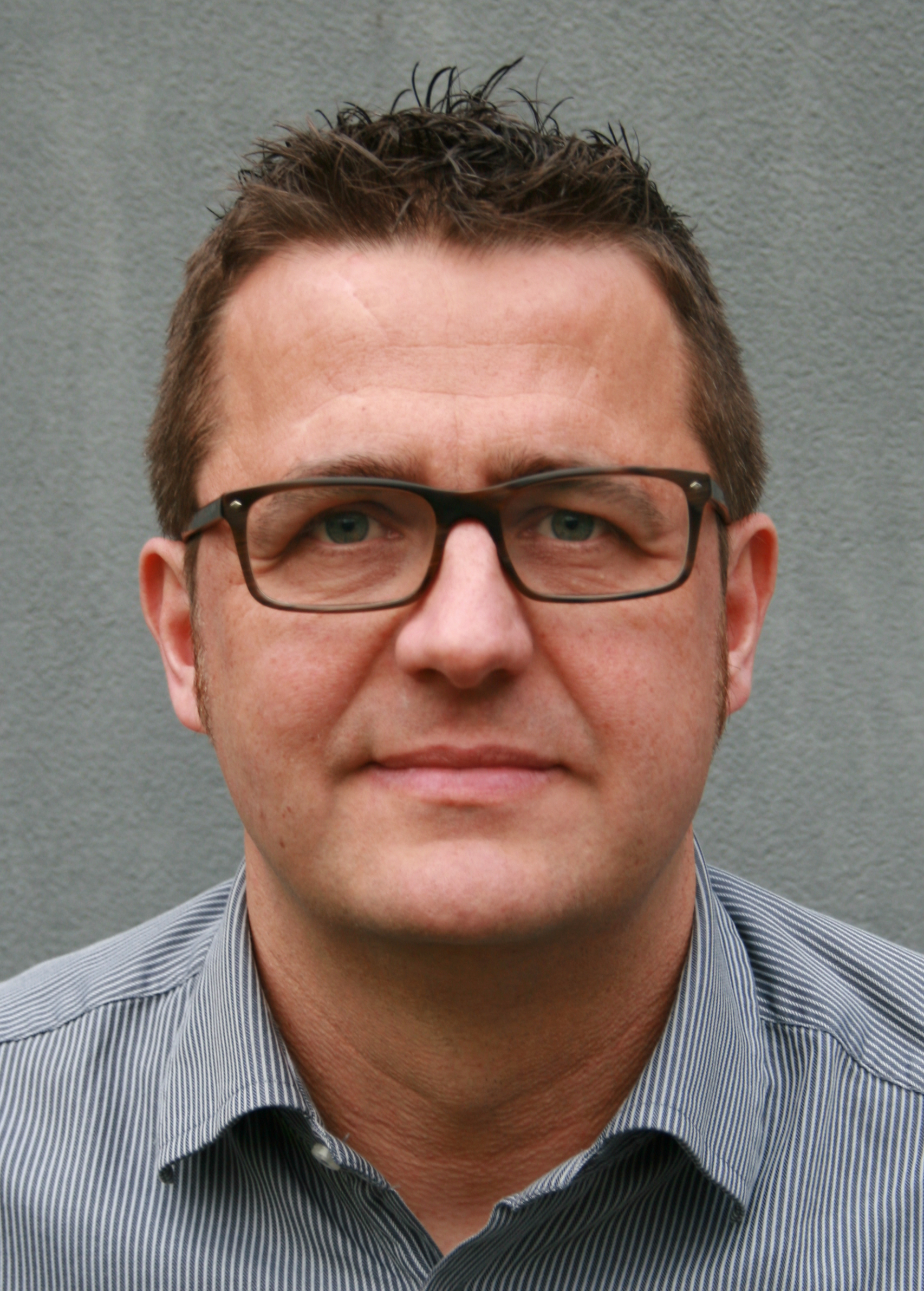}}]
{Bjorn De Sutter} is associate professor at Ghent University in the Computer
Systems Lab. He obtained his MSc and PhD degrees in Computer Science from Ghent
University's Faculty of Engineering in 1997 and 2002. His research focuses on
the use of compiler techniques to aid programmers with non-functional aspects of
their software, such as performance, code size, reliability, and security.
\end{IEEEbiography}
\vfill
\end{document}